\documentstyle[graphicx,prc,aps,floats,preprint,tighten]{revtex}

\newif\iffloat\floattrue

\newcommand{\gold}{{}^{197}\!{\mathrm{Au}}}
\newcommand{\calcium}{{}^{40}{\mathrm{Ca}}}
\newcommand{\carbon}{{}^{12}{\mathrm{C}}}
\renewcommand{\Re}{\mathop{\mathrm{Re}}}
\renewcommand{\Im}{\mathop{\mathrm{Im}}}

\begin{document}
\draft
\title{Antisymmetrized molecular dynamics with quantum
 branching processes for collisions of heavy nuclei}
\author{Akira Ono}
\address{Department of Physics, Tohoku University, Sendai 980-8578,
Japan.}
\date{\today}
\maketitle

\begin{abstract}
Antisymmetrized molecular dynamics (AMD) with quantum branching
processes is reformulated so that it can be applicable to the
collisions of heavy nuclei such as $\gold+\gold$ multifragmentation
reactions.  The quantum branching process due to the wave packet
diffusion effect is treated as a random term in a Langevin-type
equation of motion, whose numerical treatment is much easier than the
method of the previous papers.  Furthermore a new approximation
formula, called the triple-loop approximation, is introduced in order
to evaluate the Hamiltonian in the equation of motion with much less
computation time than the exact calculation.  A calculation is
performed for the $\gold+\gold$ central collisions at 150 MeV/nucleon.
The result shows that AMD almost reproduces the copious fragment
formation in this reaction.
\end{abstract}

\narrowtext
\section{Introduction}

Various kinds of microscopic dynamical models have been developed in
order to understand the various phenomena in heavy ion collisions in
the medium energy region. The mean field models, such as the
time-dependent Hartree-Fock (TDHF) theory and the
Vlasov-Uehling-Uhlenbeck (VUU) equation {}\cite{BERTSCH,CASSING}, are
good at the precise description of the single particle dynamics in the
mean field.  On the other hand, the advantage of the molecular
dynamics models~\cite{AICHELIN,MARUb,ONOab,FELDMEIER} is, generally
speaking, that they can describe the many-body correlation which is
essential in the fragment formation.

Nuclear multifragmentation has been a hot topics in these years.  It
can be regarded as a manifestation of the liquid-gas phase transition
in the nuclear matter, and we expect that the precious information of
the finite-temperature nuclear matter in high and low density can be
obtained by studying the multifragmentation reactions.  Although the
multifragmentation should be related to the property of nuclear
matter, the ideally equilibrated nuclear matter is not formed in the
real reactions, and therefore the studies by using the microscopic
dynamical models are indispensable.  Furthermore the
multifragmentation is a good touchstone for the microscopic models
because it includes the nontrivial mechanism for the fragment
formation, namely the appearance of new cluster correlations with
dynamical symmetry breaking from an almost uniform excited matter.

From the viewpoint of the time-dependent quantum theories which solve
the time evolution of the system from the given initial state to the
final state, the multifragmentation is not easy to treat, because the
final state should be a superposition of a huge number of the channel
wave functions.  There are a huge number of possible ways to decompose
the total system into fragments.  The initial state of the reaction
and the individual channels of the intermediate and final states may
be well described by using rather simple wave functions, but the total
wave function of the intermediate or final state is, of course, too
much complicated to handle [See Fig.\ {}\ref{fig:multichannel}].
However, since the interference among channels is not so important in
usual cases, the quantum branchings from a single channel to the
superposition of many channels can be treated as stochastic branching
processes without taking account of the interference among channels.
Namely, in a practical time-dependent model where each channel is
described by a rather simple wave function, the time evolution of the
system should be determined by the successive stochastic quantum
branching processes in addition to the deterministic time evolution
within each channel.  The necessary quantum branching process varies
according to the model because it should depend on how the channel
wave function is restricted to the simple one.  The physical
observables are calculated as the ensemble average values of the
expectation values all over the channels.

\iffloat
\begin{figure}
\begin{center}
\begin{minipage}{0.45\textwidth}
\begin{center}
\includegraphics[width=\textwidth]{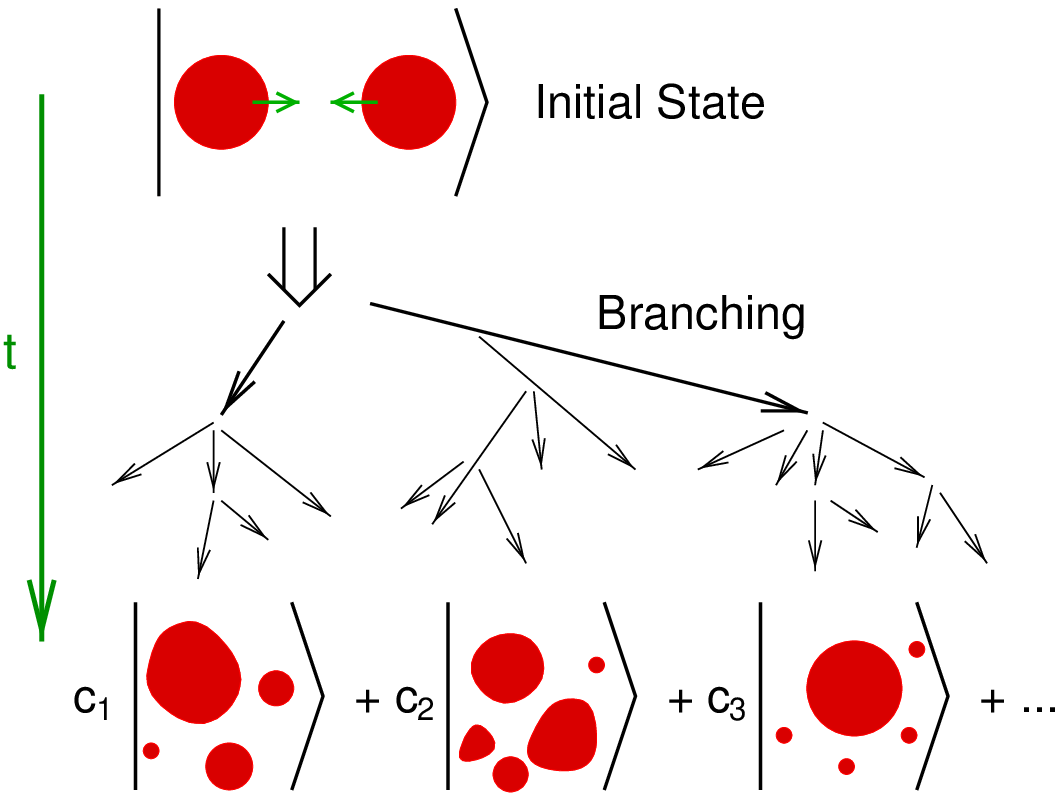}
\end{center}
\end{minipage}
\end{center}
\caption{\label{fig:multichannel}
A schematic picture of the quantum branching processes for
multichannel reactions.}
\end{figure}
\fi

As is well known, TDHF is not suitable for the reactions with many
channels such as multifragmentation because it does not take account
of the quantum branchings mentioned above.  Although a single Slater
determinant may be sufficient for the initial two nuclei and the
fragments in each final channel, it is far from sufficient for the
superposition of the final channel wave functions.  In such cases,
what one can expect by solving the deterministic time evolution is, at
best, that one of the possible channels appears as the final state.
In bad cases, however, the obtained final state looks like none of the
final channels.  The latter may be the case for the TDHF application
to the multifragmentation of an expanding system because the mean field
gets weaker and weaker as the system expands widely and therefore the
diffusing single particle wave functions will never relocalize to form
clusters.  It is dangerous to apply the mean field concept to dilute
system where the system may be branching into channels whose mean
fields should be different from channel to channel.  The VUU equation
can be regarded as one of the extensions of TDHF, where the two-body
collision effect is introduced as a term in the equation of the
one-body phase space distribution.  However, since this collision term
is a deterministic term to take account of the averaged effect of the
two-nucleon collisions, the quantum branching effect is not included
in the VUU equation.  The Boltzmann Langevin approach
{}\cite{AYIK,RANDRUP}, on the other hand, introduced the fluctuation
as a random term associated with the two-nucleon collisions, and it
may be applicable to the multichannel reactions such as
multifragmentation.  However, it is difficult to understand that the
real origin of the cluster formation were the fluctuation due to the
two-nucleon collisions, because the two-nucleon collisions are rare in
the expanding nuclear matter, while the cluster formation should take
place even in the ideal situation where the nuclear matter is
uniformly expanding without initial fluctuation.

On the contrary to the mean field models mentioned above, the
molecular dynamics models restrict the channel wave function to an
(antisymmetrized) product of wave packets.  The shape of the wave
packets are usually kept fixed and the many body wave function is
parametrized only by the centroids of the wave packets.  The benefit
of this restriction is that we can avoid encountering the situation
where the single particle wave functions have been expanded and then
the mean field concept does not work any longer.  In other words, the
channel wave function in the molecular dynamics models are restricted
so that it cannot be a mixture of many channels which should be
treated independently rather than as a whole in a single wave
function.  In the quantum molecular dynamics (QMD)
{}\cite{AICHELIN,MARUb} and the antisymmetrized molecular dynamics
(AMD) {}\cite{ONOab,FELDMEIER}, the centroid motion in each channel is
determined by the equation of motion derived from the time-dependent
variational principle.  In addition to it, the effect of the two-body
collisions is introduced as a stochastic branching process, which
brings the system from a single initial state to many possible final
channels randomly.  Because of these reasons, QMD and AMD are suitable
frameworks for the fragmentation phenomena.

Many people have ever tried to extend the molecular dynamics models by
generalizing the wave packets
{}\cite{KIDERLEN,CHOMAZ,MARU-EQMD,FELDMEIER}, usually by treating the
width parameters of the wave packets as dynamical variables as well as
the wave packet centroids.  Although this extension can be an
improvement for some phenomena {}\cite{MARU-EQMD}, it is rather a
change for the worse in the context of multifragmentation.  Such
extension draws the molecular dynamics models close to a mean field
model which fails into the pathological situation where many
independent channels are mixed in a single Slater determinant.  In
fact, Kiderlen et al.\ {}\cite{KIDERLEN} and Chomaz et al.\
{}\cite{CHOMAZ} reported that the diffused wave packets never shrink
again to form clusters in a hot expanding nuclear system.

In Ref.\ {}\cite{ONOh}, we took a different way to extend AMD by
taking account of the precise one-body dynamics without losing the
benefit of the molecular dynamics models that the channel wave
function cannot fall into a mixture of many channels.  It was achieved
not by generalizing the channel wave function but by introducing the
wave packet diffusion effect as a new quantum branching process.  This
extended AMD is called AMD-V, since the wave packet diffusion effect
is calculated with the Vlasov equation {}\cite{WONG}.  When a
expanding system is calculated by AMD-V, one can imagine that not only
the centroids expand but also the successive quantum branchings take
place due to the wave packet diffusion effect, and that the
multifragment channels appear stochastically.  In fact, we showed in
Refs.\ {}\cite{ONOh,WADA} that AMD-V works very well for the
multifragmentation in $\calcium+\calcium$ reaction at 35 MeV/nucleon,
though the expansion in this case is not spherical but mainly in the
beam direction like the neck fragmentation.  The wave packet diffusion
process is also related to the nucleon emission rate and the energy
carried out by emitted nucleons, which was essential for the correct
prediction of the excitation energies of the produced fragments.  No
other microscopic models have ever reproduced these fragmentation data
so nicely.

Ohnishi and Randrup take yet another approach to improve the molecular
dynamics models {}\cite{OHNISHI-RANDRUP}.  Based on the idea that the
essential part of the multifragmentation is governed by the
statistical effect, they introduced a fluctuation-dissipation term to
the equation of motion by hand so as to ensure the correct equilibrium
property.  Although the good statistics is the advantage of their
model, there is no microscopic and dynamical background for the added
fluctuation-dissipation term.  We would like to emphasize here that
their approach is not the unique way to get the quantum statistics in
the molecular dynamics.  Even though we start with the microscopic
dynamical consideration, it is possible to get the quantum statistics
as shown in Refs. {}\cite{ONOf,ONOg}.

In spite of the fact that interesting high-quality multifragmentation
data were published for heavy system such as $\gold+\gold$ collisions
{}\cite{REISDORF}, no satisfactory explanation by microscopic
dynamical models has been given.  This difficulty is due to the
essentially quantum mechanical feature of multifragmentation.
Although AMD-V is one of few realistic models that have possibility to
reproduce the data, it was impossible so far to apply AMD to heavy
systems because of the reason of CPU time.  The main purpose of this
paper is, therefore, to give a framework of AMD-V whose numerical
calculation is feasible even for $\gold+\gold$ collisions, by
introducing an improvement and an approximation to the original AMD-V
framework.

The necessary CPU time of the original AMD calculation is proportional
to the forth power of the mass number of the system. This means that
the required CPU time for $\gold+\gold$ reactions is about 600 times
as much as for $\calcium+\calcium$ reactions.  In order to overcome
this problem, we introduce in this paper a new approximation for the
AMD Hamiltonian which can be evaluated with the CPU time proportional
to the third power of the mass number.  This approximation is called
the triple-loop approximation.

In the original AMD-V calculation of Ref.\ {}\cite{ONOh}, the most
time-consuming part was the procedure to ensure the energy
conservation after the quantum branching process due to the wave
packet diffusion effect.  It was necessary to solve a kind of
frictional cooling equation at least for several time steps to search
the energy conserving point.  This procedure becomes unnecessary and
the framework becomes more transparent in this paper when the wave
packet diffusion effect is reformulated as a random term in a
Langevin-type equation of motion, the formal structure of which is
similar to the equation by Ohnishi and Randrup
{}\cite{OHNISHI-RANDRUP}.

This paper is organized as follows.  In Sec.\ II, the framework of the
improved AMD-V is given.  Especially the wave packet diffusion process
is formulated as a random term in the equation of motion.  In Sec.\
III, the triple-loop approximation for the AMD Hamiltonian is
formulated and some tests for this approximation are given.  A
demonstrative calculation for $\gold+\gold$ collisions at 150
MeV/nucleon is given in Sec.\ IV in order to show that the AMD-V
calculation for heavy system is really possible and it is likely to
reproduce the multifragmentation data.  Section V is devoted for the
summary.

\section{Framework of AMD with Quantum Branching Processes}

In constructing a time-dependent quantum model for medium energy heavy
ion collisions, one should keep in mind the fact that the initial
state branches into a huge number of reaction channels in the
intermediate states and the final state.  It is too difficult to
follow the time evolution of the total many-body wave function in
which the many-body correlations are not negligible.  Therefore, we
treat separately the branching into channels and the time evolution
within each channel.  Approximations such as the mean field theory may
be valid within each channel, while the interference among the
branched channels may be unimportant.  The independence of the time
evolution of each channel should be respected.

\subsection{Channel wave function and equation of motion}

We describe each channel wave function by an AMD wave function which
is a single Slater determinant of Gaussian wave packets
{}\cite{ONOab},
\begin{equation}
\Phi(Z)=
\det\Bigl[\exp\Bigl\{
  -\nu\Bigl({\mathbf{r}}_j - \frac{{\mathbf{Z}}_i}{\sqrt\nu}\Bigr)^2
  +\frac{1}{2}{\mathbf{Z}}_i^2\Bigr\}
\chi_{\alpha_i}(j)\Bigr], 
\label{eq:AMDWaveFunction}
\end{equation}
where the complex variables $Z\equiv\{{\mathbf{Z}}_i;\
i=1,\ldots,A\}=\{Z_{i\sigma};\ i=1,\ldots,A,\ \sigma=x,y,z\}$
represent the centroids of the wave packets.  We take the width
parameter $\nu=0.16$ $\mathrm{fm}^{-2}$ and the spin isospin states
$\chi_{\alpha_i}=p\uparrow$, $p\downarrow$, $n\uparrow$, or
$n\downarrow$.  

The AMD wave function (\ref{eq:AMDWaveFunction}) seems to be very
simple, but it is sufficient for the description of the ground states
of nuclei.  For example, the binding energies obtained by the
frictional cooling method {}\cite{KANADAENYOa} coincide with the
experimental data within the precision of 1 MeV/nucleon even though
the common values of $\nu$ and $T_0$ (mentioned later) are used for
all nuclei {}\cite{ONOd}. Therefore the initial state of the reaction
and the individual channel wave functions in the intermediate and
final states are well described by the AMD wave functions.

The time evolution of the wave packet centroids $Z$ within the same
channel is determined by the time dependent variational principle,
\begin{equation}
\delta\int dt\, 
  {\langle\Phi(Z)|(i\hbar{d\over dt}-H)|\Phi(Z)\rangle
  \over\langle\Phi(Z)|\Phi(Z)\rangle}=0,
\end{equation}
from which one can derive the equation of motion for $Z$,
\begin{equation}
  i\hbar\sum_{j\tau}C_{i\sigma,j\tau}{dZ_{j\tau}\over dt}=
  {\partial{\mathcal{H}}\over\partial Z_{i\sigma}^*}.
  \label{eq:AMDEqOfMotion}
\end{equation}
$C_{i\sigma,j\tau}$ with $\sigma,\tau=x,y,z$ is a hermitian matrix
defined by
\begin{equation}
C_{i\sigma,j\tau}=
\frac{\partial^2}{\partial Z_{i\sigma}^*\partial Z_{j\tau}}
\log\langle\Phi(Z)|\Phi(Z)\rangle,
\end{equation}
and $\cal H$ is the expectation value of the Hamiltonian after the
subtraction of the spurious kinetic energy of the zero-point
oscillation of the center-of-masses of fragments {}\cite{ONOab},
\begin{equation}
  {\cal H}(Z)=\frac{\langle\Phi(Z)|H|\Phi(Z)\rangle}
                    {\langle\Phi(Z)|\Phi(Z)\rangle}
          -\frac{3\hbar^2\nu}{2M}A+T_0(A-N_{\mathrm{F}}(Z)).
  \label{eq:AMDHamil}
\end{equation}
The quantum Hamiltonian
\begin{equation}
H=\sum_{i=1}^{A}\frac{{\mathbf{p}}_i^2}{2M}+\sum_{i<j}v_{ij}
\end{equation}
includes an effective two-body interaction such as the Gogny force
{}\cite{GOGNY} which can be density dependent, since the model wave
function is limited to a single Slater determinant.  The spurious
kinetic energies of the zero-point oscillation of the center-of-mass
of the isolated fragments and nucleons have been subtracted in Eq.\
(\ref{eq:AMDHamil}) by introducing a continuous number of fragments
$N_{\mathrm{F}}(Z)$ {}\cite{ONOab}.  Without this subtraction, the
Q-values for nucleon emissions and fragmentations would not be
reproduced.  The parameter $T_0$ is $3\hbar^2\nu/2M$ in principle but
treated as a free parameter for the adjustment of the binding
energies.  Once the zero-point kinetic energies have been subtracted,
the center-of-mass wave function of an isolated fragment (or nucleon)
should be regarded as a plane wave.  This method, however, only takes
account of the expectation value of the kinetic energy and ignores its
quantum fluctuation.  This point will be reconsidered in the next
subsection.

For the later formulation, it is convenient to introduce the Poisson
bracket $\{{\mathcal{F}},{\mathcal{G}}\}$ and the inner product of the
canonical gradients $({\mathcal{F}},{\mathcal{G}})$,
\begin{eqnarray}
\{{\mathcal{F}},{\mathcal{G}}\}&=&
\frac{1}{i\hbar}\sum_{i\sigma,j\tau}
  \frac{\partial{\mathcal{F}}}{\partial Z_{i\sigma}}
  C^{-1}_{i\sigma,j\tau}
  \frac{\partial{\mathcal{G}}}{\partial Z_{j\tau}^*}\nonumber\\
&&-\frac{1}{i\hbar}\sum_{i\sigma,j\tau}
  \frac{\partial{\mathcal{G}}}{\partial Z_{i\sigma}}
  C^{-1}_{i\sigma,j\tau}
  \frac{\partial{\mathcal{F}}}{\partial Z_{j\tau}^*},
\label{eq:PoissonBracket}
\\
({\mathcal{F}},{\mathcal{G}})&=&
\frac{1}{\hbar}\sum_{i\sigma,j\tau}
  \frac{\partial{\mathcal{F}}}{\partial Z_{i\sigma}}
  C^{-1}_{i\sigma,j\tau}
  \frac{\partial{\mathcal{G}}}{\partial Z_{j\tau}^*}\nonumber\\
&&+\frac{1}{\hbar}\sum_{i\sigma,j\tau}
  \frac{\partial{\mathcal{G}}}{\partial Z_{i\sigma}}
  C^{-1}_{i\sigma,j\tau}
  \frac{\partial{\mathcal{F}}}{\partial Z_{j\tau}^*}.
\end{eqnarray}
Then the equation of motion (\ref{eq:AMDEqOfMotion}) can be rewritten
as
\begin{equation}
\dot{Z}=\{Z,{\mathcal{H}}\}.
\end{equation}
On the other hand, the frictional cooling equation can be written as
\begin{equation}
\dot{Z}=\lambda\{Z,{\mathcal{H}}\}+\mu(Z,{\mathcal{H}}),
\label{eq:EqOfCooling}
\end{equation}
for which one can show that the energy expectation value decreases as
time
\begin{equation}
\dot{\mathcal{H}}=\mu({\mathcal{H}},{\mathcal{H}})\le0
\end{equation}
for arbitrary $\lambda$ and $\mu(<0)$.

The one-body Wigner function for the AMD wave function
(\ref{eq:AMDWaveFunction}) is given by
\begin{eqnarray}
f({\mathbf{r}},{\mathbf{p}})&=&8\sum_{ik}
e^{-2({\mathbf{u}}^*-{\mathbf{Z}}_i^*)\cdot({\mathbf{u}}-{\mathbf{Z}}_k)}
B_{ik}B^{-1}_{ki}
\label{eq:AMDWignerFunction}
,\\
{\mathbf{u}}&=&\sqrt{\nu}\,{\mathbf{r}}+\frac{i}{2\hbar\sqrt{\nu}}{\mathbf{p}},
\end{eqnarray}
where
\begin{equation}
B_{ik}=e^{{\mathbf{Z}}_i^*\cdot{\mathbf{Z}}_k}\delta_{\alpha_i\alpha_k}
\end{equation}
is the overlap matrix of the non-orthogonal single particle wave
packets in Eq.\ (\ref{eq:AMDWaveFunction}).  In what follows, it is
sometimes convenient to introduce a QMD-like approximation,
\begin{equation}
f({\mathbf{r}},{\mathbf{p}})\approx 8\sum_{i}
e^{-2|{\mathbf{u}}-{\mathbf{W}}_i|^2},
\label{eq:WignerFunctionQMD}
\end{equation}
by using the physical coordinate $W=\{{\mathbf{W}}_i\}$ {}\cite{ONOab}
defined by
\begin{equation}
  {\mathbf{W}}_i=\sum_{j=1}^A \Bigl(\sqrt Q\Bigr)_{ij}{\mathbf{Z}}_j,\quad
  Q_{ij} =B_{ij}B^{-1}_{ji}.
  \label{eq:PhysCoord}
\end{equation}
The coordinates $W$ can be regarded as physical ones because
quantities such as the orbital angular momentum and the number of the
harmonic-oscillator quanta are written in the usual way by using $W$.
Furthermore the coordinates $W$ are canonical coordinates when the
antisymmetrization among more than two packets is negligible.  The
physical coordinates are very useful in various places of the
formulation of AMD-V, while the QMD-like approximation of Eq.\
(\ref{eq:WignerFunctionQMD}) is too poor to be useful in the
evaluation of the Hamiltonian as will be seen in Sec.\ III.

\subsection{Wave packet diffusion process}

The wave packet shape is not allowed to change in AMD. Therefore the
dynamics of the single particle wave functions is not so precisely
described as in TDHF.  However, we should not extend AMD to TDHF
because TDHF has the pathological problem of the spurious coupling of
channels.  In Ref.\ {}\cite{ONOh}, instead of extending the channel
wave functions, we introduced the precise single particle dynamics
into AMD as a new stochastic branching process.  In this subsection,
this process is reformulated as a random term of a Langevin-type
equation of motion which is more suitable for numerical calculations.

\subsubsection{Fluctuation due to the wave packet diffusion}

Each nucleon $k$ in an AMD wave function $\Phi(Z(t))$ of one of the
branches at the time $t$ is represented approximately by a Gaussian
wave packet in phase space
\begin{equation}
f_k(x,t)=8\, \exp\Bigl[-2\sum_{a=1}^6 (x_a-X_{ka}(t))^2\Bigr],
\end{equation}
where we have introduced the 6-dimensional phase space coordinate
\begin{equation}
\{x_a;\ a=1,\ldots,6\}=\{\sqrt\nu\,{\mathbf{r}},
{\mathbf{p}}/2\hbar\sqrt\nu\}.
\end{equation}
The centroid $\{X_{ka};\ a=1,\ldots,6\}$ stands for the physical
coordinate ${\mathbf{W}}_k$. In the usual AMD, the time evolution of
$X_{ka}$ is derived from the equation of motion while the shape of
the wave packet is fixed.

However, more reliable time evolution of the one-body distribution
function is given by the TDHF equation or the Vlasov equation
{}\cite{WONG},
\begin{equation}
{\partial f_k\over\partial t}
+{\partial h\over\partial{\mathbf{p}}}
  \cdot{\partial f_k\over\partial{\mathbf{r}}}
-{\partial h\over\partial{\mathbf{r}}}
  \cdot{\partial f_k\over\partial{\mathbf{p}}}
=0.
\label{eq:VlasovEquation}
\end{equation}
Writing the expectation value of the Hamiltonian as ${\mathcal{H}}[f]$
for a Slater determinant represented by
$f({\mathbf{r}},{\mathbf{p}})$, one can obtain the single-particle
Hamiltonian $h$ by
\begin{equation}
h({\mathbf{r}},{\mathbf{p}},t)=
\frac{\delta{\mathcal{H}}[f]}{\delta f({\mathbf{r}},{\mathbf{p}})}
\bigg|_{f=f({\mathbf{r}},{\mathbf{p}},t)},
\end{equation}
for the AMD wave function $\Phi(Z(t))$ whose Wigner function is given
by Eq.\ (\ref{eq:AMDWignerFunction}).  The time derivative of the
width and shape of the wave packet
\begin{equation}
\dot\sigma^2_{kab}(t) \equiv
\frac{d}{dt}\int \Bigl(x_a-X_{ka}(t)\Bigr)
                       \Bigl(x_b-X_{kb}(t)\Bigr) f_k(x,t) d^6x
\end{equation}
can be evaluated based on the Vlasov equation
(\ref{eq:VlasovEquation}) by using the test particle method or by the
direct analytical calculation.

It will be useful to note that the wave packet diffusion
$\dot{\sigma}^2_{kab}$ is mainly determined by the curvature of the
mean field in $h({\mathbf{r}},{\mathbf{p}})$ in the region around the
wave packet $k$.  When the potential is quadratic with the curvature
$\frac{1}{2}m\omega^2=2\hbar^2\nu^2/m$, the wave packet diffusion
effect is exactly zero, which is approximately satisfied for the
packets inside the nucleus.  On the other hand, for the packets near
the surface of the nucleus, the potential curvature is negative and
then the wave packet diffusion effect becomes essential.

Instead of changing the shape of the wave packet $f_k$, we now give
fluctuation $\delta X_{ka}(t)$ to the centroid $X_{ka}(t)$ in order to
introduce the wave packet diffusion effect $\dot\sigma^2_k$. If we
assume the white noise for the fluctuation, it should satisfy
\begin{eqnarray}
\overline{\delta X_{ka}(t)}&=&0,\\
\overline{\delta X_{ka}(t)\delta X_{kb}(t')}
&=&[\dot\sigma^2_k]_{ab}(t)\delta(t-t').
\label{eq:FlctVariance}
\end{eqnarray}
The negative eigenvalues of $\dot\sigma^2_k$ have been replaced by
zero because the shrinking of the wave packet cannot be treated unless
we respect the interference among channels.

Although the higher moments of the fluctuation can also be calculated
with the Vlasov equation, we expect that their effect is not
important. In our early work {}\cite{ONOh}, we took the distribution
function for the fluctuation $\xi_a\equiv\delta X_{ka}$,
\begin{eqnarray}
P(\xi)&=&(1-c)\delta(\xi)+c 
   \frac{\sqrt{\det\alpha}}{(\pi/2)^3}
   \exp\Bigl(-2\sum_{ab}\xi_a\alpha_{ab}\xi_b\Bigr),\\
  &&\alpha_{ab}\equiv\frac{\mathop{\mathrm{tr}}[\dot\sigma^2_k]}{3}
              [\dot\sigma^2_k]^{-1}_{ab},
\end{eqnarray}
where $c$ is chosen to give the correct variance of the fluctuation
[Eq.\ (\ref{eq:FlctVariance})]. Since $\alpha$ is of the order of 1, a
big branching takes place with small probability, while no branching
takes place in most cases. However, we here take the Gaussian
distribution
\begin{eqnarray}
P(\xi)&=&
   \frac{\sqrt{\det\alpha'}}{(\pi/2)^3}
   \exp\Bigl(-2\sum_{ab}\xi_a\alpha'_{ab}\xi_b\Bigr),\\
 &&\alpha'_{ab}\equiv c'[\dot\sigma^2_k]^{-1}_{ab},
\end{eqnarray}
where $c'$ is determined by Eq.\ (\ref{eq:FlctVariance}),
because this is more convenient for the numerical calculation. In this
case, a small fluctuation is given to each centroid at every time
step.

It should be noted that the fluctuation $\delta X_{ka}(t)$ is spurious
for an isolated wave packet $k$ because there is no other packets that
can absorb the recoil from the fluctuation.  Furthermore the mean
field theory [Eq.\ (\ref{eq:VlasovEquation})] is not necessarily valid
for light nuclei with $A\lesssim10$.  We should avoid the situation
where the unreliable fluctuation for the packets inside a light
fragment has a drastic effect on the dynamics such as spuriously
breaking the fragment.  Therefore, by checking the packets in the
neighborhood of the packet $k$, we put $\delta X_{ka}(t)=0$ when
\begin{mathletters}
\begin{equation}
\sum_i\theta\Bigl(1.75-|\Re({\mathbf{Z}}_i-{\mathbf{Z}}_k)|\Bigr)\le10
\end{equation}
and
\begin{equation}
\biggl|\sum_i\theta\Bigl(1.75-|\Re({\mathbf{Z}}_i-{\mathbf{Z}}_k)|\Bigr)
       \Re({\mathbf{Z}}_i-{\mathbf{Z}}_k)\biggr|\le5.
\end{equation}
\end{mathletters}
Although this prescription may make the cooling of light fragments too
slow, it is not a problem practically because the decay of these
fragments can be calculated later by a statistical decay code.

For numerical convenience, we now introduce a small delay time $\tau$
of the response to the fluctuation $\delta X_{ka}$.  The delayed
fluctuation $\Xi_{ka}$ is obtained by the equation
\begin{equation}
\frac{d}{dt}\Xi_{ka}(t)=\frac{1}{\tau}\delta X_{ka}(t)
                       -\frac{1}{\tau}\Xi_{ka}(t),
\label{eq:EqOfXi}
\end{equation}
whose solution is
\begin{equation}
\Xi_{ka}(t)=
\frac{1}{\tau}\int_{0}^{t} \delta X_{ka}(t')e^{-(t-t')/\tau}dt',
\end{equation}
by assuming $\Xi_{ka}(0)=0$ for the initial state.  Instead of the
original fluctuation $\delta X_{ka}(t)$, this delayed fluctuation
$\Xi_{ka}(t)$ is to be added to the centroid $X_{ka}(t)$.  In
numerical calculations we take $\tau=5$ fm/$c$, which should be
smaller than the important time scales of the reaction.  Since the
fluctuation is smoothened by the averaging over the time $\tau$, it
can be treated easily numerically.

Some readers may be interested in the difference between the
fluctuation introduced by Ohnishi and Randrup {}\cite{OHNISHI-RANDRUP}
and that of our present work.  In our model, the fluctuations of
different packets (labeled by $k$) are independent while the
correlations of the phase space components (labeled by $a$ and $b$) of
each packet is properly incorporated by Eq.\ (\ref{eq:FlctVariance}).
This is a natural consequence of the fact that our fluctuation is
introduced based on the mean field model.  On the contrary, Ohnishi
and Randrup simply ignores the importance of the phase space
correlations, while they introduce the correlations among different
packets without any microscopic or dynamical justification.

\subsubsection{Equation of motion and conserved quantities}

The above determined fluctuation $\Xi_{ka}(t)$ or its complex vector
representation $\bbox{\Xi}_{k}(t)$ is the fluctuation to the physical
coordinate ${\mathbf{W}}_k$.  In order to put it in the equation of
motion, it is now necessary to convert it to the fluctuation to the
original AMD coordinates $Z$.  For this purpose let us introduce a
time-dependent one-body hermitian operator $\hat{o}_k(t)$ that
generates the fluctuation $\bbox{\Xi}_{k}(t)$.  The form of
$\hat{o}_k(t)$ is taken as
\begin{eqnarray}
\hat{o}_k(t)=i\sum_{j=1}^{A}\Bigl\{
&&  ({\mathbf{y}}_{kj}(t)\cdot\hat{{\mathbf{a}}}^\dagger)
  |{\mathbf{W}}_j(t)\rangle\langle{\mathbf{W}}_j(t)| \nonumber\\
&&- |{\mathbf{W}}_j(t)\rangle\langle{\mathbf{W}}_j(t)|
  ({\mathbf{y}}_{kj}^*(t)\cdot\hat{{\mathbf{a}}}) \Bigr\},
\end{eqnarray}
where the stochastic complex parameters $\{{\mathbf{y}}_{kj}(t);\
j=1,\ldots,A\}$ are to be determined below, and
\begin{eqnarray}
\hat{\mathbf{a}}&=&\sqrt{\nu}\,\hat{\mathbf{r}}
                +\frac{i}{2\hbar\sqrt{\nu}}\hat{\mathbf{p}},\\
\langle{\mathbf{r}}|{\mathbf{W}}\rangle
&\propto& \exp\Bigl\{
         -\nu\Bigl({\mathbf{r}}-\frac{\mathbf{W}}{\sqrt{\nu}}\Bigr)^2\Bigr\}.
\end{eqnarray}
In QMD-like approximation by the use of the physical coordinates, the
expectation value of this one-body operator is calculated as
\begin{eqnarray}
&&{\mathcal{O}}'_k(W,t)
=\sum_{i=1}^{A}\langle{\mathbf{W}}_i|\hat{o}_k|{\mathbf{W}}_i\rangle\\
&&\quad=i\sum_{ij}({\mathbf{y}}_{kj}(t)\cdot{\mathbf{W}}_i^*
           -{\mathbf{y}}_{kj}^*(t)\cdot{\mathbf{W}}_i)
           e^{-|{\mathbf{W}}_j(t)-{\mathbf{W}}_i|^2}
\end{eqnarray}
By identifying the physical coordinates $W$ with the canonical
coordinates, the stochastic parameters $\{{\mathbf{y}}_{kj}(t)\}$ are
determined by the requirement that the one-body operator generates the
fluctuation $\bbox{\Xi}_k(t)$ at the moment $t$,
\begin{equation}
i\hbar\delta_{ik}\bbox{\Xi}_{k}(t)
=\frac{\partial{\mathcal{O}}'_k(t)}{\partial{\mathbf{W}}_i^*}
                 \biggl|_{W=W(t)}.
\end{equation}
Then the fluctuation for $Z$ should be generated as
$\{Z,{\mathcal{O}}_k(t)\}$ by the exact expectation value of the same
one-body operator,
\begin{eqnarray}
{\mathcal{O}}_k(Z,t)&=&\frac{\langle\Phi(Z)|\hat{O}_k(t)|\Phi(Z)\rangle}
                  {\langle\Phi(Z)|\Phi(Z)\rangle},\\
\hat{O}_k(t)&=&\sum_{i=1}^{A}\hat{o}_{ki}(t).
\end{eqnarray}

Before putting the fluctuation $\{Z,{\mathcal{O}}_k(t)\}$ in the
equation of motion, we should note the fact that the fluctuation
violates the conservation lows for the total momentum and the total
energy.  Such conservation lows should be achieved through many-body
correlations in reality.  Since this kind of many-body correlations
are beyond the scope of the one-body dynamics of the Vlasov equation,
it is inevitable to introduce the conservation lows by hand.  By
correcting the fluctuation for the conservation lows, the equation of
motion of AMD-V is now written as
\begin{eqnarray}
\dot{Z}=\{Z,{\mathcal{H}}\}
+\sum_{k=1}^{A}\gamma_k\biggl[
\Bigl\{Z,\;{\mathcal{O}}_k
        +\sum_m\alpha_{km}{\mathcal{P}}_m\Bigr\}_{{\mathrm{C}}_k}
&&\nonumber\\
+\mu_k\Bigl(Z,\;{\mathcal{H}}
        +\sum_m\beta_{km}{\mathcal{Q}}_m\Bigr)_{{\mathrm{N}}_k}
&&\biggr].
\label{eq:AMDVEqOfMotion}
\end{eqnarray}
The first term in the square bracket is the fluctuation due to
$\bbox{\Xi}_k$ corrected for the center-of-mass coordinate and
momentum conservation, and the second term is the cooling (or heating)
term to ensure the energy conservation.  The parameter $\gamma_k$ can
be regarded as 1 until its meaning is explained later.

When the system has been decomposed into several clusters, the
fluctuation $\bbox{\Xi}_k$ to a packet $k$ in one of the clusters
should not affect the packets in the other clusters through the
conservation lows.  In order to ensure this point, we define the
cluster ${\mathrm{C}}_k$ which includes the packet $k$, where the
clusters are identified by the condition that two packets $i$ and $j$
belong to the same cluster if $|{\mathbf{Z}}_i-{\mathbf{Z}}_j|<1.75$.
The subscript ${\mathrm{C}}_k$ of the Poisson bracket in Eq.\
(\ref{eq:AMDVEqOfMotion}) indicates that the centroids of the packets
in the other clusters are treated as static parameters.  Namely, the
packets in the other clusters are omitted in the summation in Eq.\
(\ref{eq:PoissonBracket}), and $C^{-1}$ is replaced by the inverse
matrix of the submatrix of $C$.  In Eq.\ (\ref{eq:AMDVEqOfMotion}), by
using the Lagrange multipliers $\alpha_m$, the constraints are
introduced for the conserved quantities $\{{\mathcal{P}}_m\}$, which
are the three components of the center-of-mass coordinate and the
three components of the total momentum
\begin{eqnarray}
\bigl<\frac{1}{A}\sum_i{\mathbf{r}}_i\bigr>
&=&\frac{1}{A}\sum_i\Re{\mathbf{Z}}_i/\sqrt{\nu},\\
\bigl<\sum_i{\mathbf{p}}_i\bigr>&=&\sum_i 2\hbar\sqrt{\nu}\Im{\mathbf{Z}}_i.
\end{eqnarray}
Then the Lagrange multipliers should be determined by
\begin{equation}
\{{\mathcal{P}}_l,{\mathcal{O}}_k\}_{{\mathrm{C}}_k}
+\sum_m \{{\mathcal{P}}_l,{\mathcal{P}}_m\}_{{\mathrm{C}}_k}\alpha_{km}=0.
\end{equation}

The method to ensure the energy conservation should be considered
carefully, because it has more drastic effects than the center-of-mass
conservation.  The set of the packets ${\mathrm{N}}_k$ which can be
adjusted in order to cancel the energy violation by $\bbox{\Xi}_k$ is
restricted to the neighborhood of the packet $k$ defined by
\begin{equation}
{\mathrm{N}}_k=\{i;\
              |{\mathbf{Z}}_i-{\mathbf{Z}}_k|<2.5\ \mbox{and}\ 
              i\in{\mathrm{C}}_k\ \mbox{and}\
              i\ne k\}.
\label{eq:DefOfNeighborhood}
\end{equation}
The total energy is restored by the frictional cooling term in Eq.\
(\ref{eq:AMDVEqOfMotion}) with $\mu_k$ adjusted for the conservation.
Since this cooling term should not violate the other conservation
lows, the quantities
\begin{equation}
\{{\mathcal{Q}}_m\}=\Bigl\{
\bigl<\sum_i {\mathbf{r}}_i\bigr>,\ 
\bigl<\sum_i {\mathbf{p}}_i\bigr>,\ 
\bigl<\sum_i{\mathbf{r}}_i\times{\mathbf{p}}_i\bigr>
\Bigr\}
\end{equation}
are kept constant by determining the Lagrange multipliers $\beta_{km}$
by
\begin{equation}
({\mathcal{Q}}_l,{\mathcal{H}})_{{\mathrm{N}}_k}
+\sum_m({\mathcal{Q}}_l,{\mathcal{Q}}_m)_{{\mathrm{N}}_k}\beta_{km}=0.
\end{equation}
The parameter $\mu_k$ is then determined by
\begin{equation}
\mu_k=
-\frac{\{{\mathcal{H}},\;{\mathcal{O}}_k+\sum_m\alpha_{km}{\mathcal{P}}_m\}
      _{{\mathrm{C}}_k}}
      {({\mathcal{H}},\;{\mathcal{H}}+\sum_m\beta_{km}{\mathcal{Q}}_m)
      _{{\mathrm{N}}_k}}
\label{eq:Muk}
\end{equation}
in order to conserve the total energy.

It should be noted that $\mu_k$ appear in Eq.\
(\ref{eq:AMDVEqOfMotion}) only through their summation
$\mu\equiv\sum_k\mu_k$ if the constrains are ignored for simplicity.
Since $\mu$ is an intensive quantity (which is independent of the size
of the system) and it is averaged over many independent fluctuations
${\mathcal{O}}_k$, one can replace $\mu$ with its averaged value
$\bar\mu$ which is a function of the current state $Z$.  Then the
cooling term of Eq.\ (\ref{eq:AMDVEqOfMotion}) is formally similar to
the dissipation term of the Langevin equation that Ohnishi and Randrup
proposed to introduce together with the fluctuation term
{}\cite{OHNISHI-RANDRUP}.  However, we use Eq.\ (\ref{eq:Muk})
directly without replacing $\mu_k$ with their averaged values, so that
the total energy is exactly conserved.  Furthermore, in our method, we
do not need to evaluate the second derivatives of the Hamiltonian
${\mathcal{H}}$ which would be necessary in order to directly evaluate
the averaged value $\bar\mu$.

The method of the energy conservation is the most difficult ambiguity
of this model because it is an effect beyond the mean field theories.
The above prescription, therefore, intend to achieve the energy
conservation with the least modification of the other degrees of
freedom by moving them in the direction of the canonical gradient of
the Hamiltonian.  However, as discussed in Ref.\ {}\cite{ONOh}, it
seems that the adjusted degrees of freedom should be restricted to the
thermal or single-particle ones in order to avoid the unphysical
direct energy conversion from the collective energy (such as the
incident energy of the heavy ion collision) to the single-particle
energy of the fluctuation.  For this purpose, the monopole and the
quadrupole moments in the coordinate and momentum spaces
\begin{equation}
\bigl<\sum_i {\mathbf{r}}_i{\mathbf{r}}_i\bigr>,\ 
\bigl<\sum_i {\mathbf{p}}_i{\mathbf{p}}_i\bigr>
\end{equation}
are also included in $\{{\mathcal{Q}}_m\}$ when ${\mathrm{N}}_k$ is
composed of more than 15 packets.

For an isolated packet $k$, we have put $\delta X_{ka}=0$.  However,
due to the delay time $\tau$, the delayed fluctuation $\bbox{\Xi}_k$
may not be zero and should be respected even for an isolated packet.
Therefore, when $N_k\le4$ with $N_k$ being the number of the element
of ${\mathrm{N}}_k$, we search the non-isolated wave packet $i$
($N_i>4$) that is the closest to the packet $k$, and then
${\mathrm{N}}_i\cup{\mathrm{N}}_k\cup\{i\}$ and
${\mathrm{C}}_i\cup\{k\}$ are used instead of ${\mathrm{N}}_k$ and
${\mathrm{C}}_k$, respectively, in the above formalism.

Finally we comment on the necessary correction when the system is near
the ground state. As already discussed in Ref.\ {}\cite{ONOh}, the
fluctuation is small but not exactly zero even for the ground state
because of the semiclassical nature of the Vlasov equation and the
restricted Slater determinant in AMD. Since the fluctuation should be
zero in the ground state, a reduction factor $\gamma_k$ is introduced
in Eq.\ (\ref{eq:AMDVEqOfMotion}) in order to cancel the fluctuation
only near the ground state.  By noting that the cooling term becomes
zero for the ground state, a measure of the difference from the ground
state is introduced by
\begin{equation}
D_k\equiv
\frac{6}{6N_k-N_{\mathrm{cons}}}
\Bigl({\mathcal{H}},\;{\mathcal{H}}+\sum_m\beta_{km}{\mathcal{Q}}_m\Bigr)
      _{{\mathrm{N}}_k},
\end{equation}
where $N_{\mathrm{cons}}$ denotes the number of the constrained
quantities $\{{\mathcal{Q}}_m\}$, and therefore $6N_k-N_{\mathrm{cons}}$
is the number of the free degrees of freedom for the energy
adjustment.  The reduction factor $\gamma_k$ is then taken as
\begin{eqnarray}
\gamma_k&=&\frac{1}{\sqrt{1+(\mu_k/\mu_{0k})^2}},\\
\mu_{0k}&=&\frac{1200}{6N_k-N_{\mathrm{cons}}}
      \sqrt{\frac{5\ \mathrm{fm}/c}{\tau}}
      \biggl(\frac{D_k}{0.1\ \mathrm{MeV}/(\mathrm{fm}/c)}\biggr)^3,
\end{eqnarray}
so that the coefficient for the cooling term $\gamma_k|\mu_k|$ does
not exceed the upper limit $\mu_{0k}$.  The purpose of the dependence
of $\mu_{0k}$ on $(6N_k-N_{\mathrm{cons}})$ and $\tau$ is to make this
reduction effect independent of the choice of the neighborhood
${\mathrm{N}}_k$ [Eq.\ (\ref{eq:DefOfNeighborhood})] and the delay
time $\tau$.  With this parameterization, the fluctuation is reduced
to zero in the ground state, while there is almost no reduction soon
after hard two-nucleon collisions in the example of $\gold+\gold$
collisions shown in Sec.\ IV.

\subsubsection{Energy fluctuation of emitted packets}

As already discussed, we have subtracted the zero-point kinetic
energies of isolated packets in Eq.\ (\ref{eq:AMDHamil}).  Therefore
the emitted packet $k$ should be regarded as a plane wave of the
momentum ${\mathbf{P}}_k=2\hbar\sqrt{\nu}\Im{\mathbf{Z}}_k$.  This is
convenient than treating it as a Gaussian packet with a momentum
spread because the nucleons in the final channels are usually observed
experimentally as momentum and energy eigen states and we should
ensure the momentum and the energy conservations in each cannel.

Let us consider the case where a wave packet $k$ would be emitted as a
whole with the momentum centroid value ${\mathbf{P}}_{0k}$ when the
zero-point kinetic energy were not subtracted from the Hamiltonian.
The momentum of this nucleon is
\begin{equation}
{\mathbf{p}}_k={\mathbf{P}}_{0k}+{\mathbf{q}},
\label{eq:MomentumOfNucleon}
\end{equation}
where ${\mathbf{q}}$ is a random number of the Gaussian distribution
with $\langle{\mathbf{q}}\rangle=0$ and $\langle q_\sigma
q_\tau\rangle=\hbar^2\nu\delta_{\sigma\tau}$ for $\sigma,\tau=x,y,z$.
The kinetic energy of this wave packet is then
\begin{equation}
E_k=\frac{{\mathbf{P}}_{0k}^2}{2M}
   +\frac{{\mathbf{P}}_{0k}\cdot{\mathbf{q}}}{M}
   +\frac{{\mathbf{q}}^2}{2M}.
\label{eq:EnergyOfPacket}
\end{equation}
When the expectation value of the third term
$\langle{\mathbf{q}}^2\rangle/2M=3\hbar^2\nu/2M$ is subtracted from the
Hamiltonian, the added term in Eq.\ (\ref{eq:AMDHamil}) acts as a
repulsive force to this packet.  Then it will be emitted with the
momentum ${\mathbf{P}}_k$ that satisfies
\begin{equation}
\langle E_k\rangle=\frac{{\mathbf{P}}_{k}^2}{2M}
=\frac{{\mathbf{P}}_{0k}^2}{2M}+\frac{3\hbar^2\nu}{2M}.
\end{equation}
Namely the momentum ${\mathbf{P}}_k$ is larger than the true centroid
${\mathbf{P}}_{0k}$ while the energy expectation value does not change
because of the absence of the momentum spread when the zero-point
energy is subtracted.

This prescription, however, has a shortcoming that it takes account of
only the expectation value of the kinetic energy and ignores its
fluctuation.  For a preequilibrium nucleon in high energy collisions,
${\mathbf{P}}_{0k}$ may be so large that the fluctuation of the second
term of Eq.\ (\ref{eq:EnergyOfPacket}) may play some role though its
expectation value is zero.  In order to take account of this kind of
energy fluctuation, we now introduce a random process when each packet
is emitted.  By neglecting the difference of the direction of
${\mathbf{P}}_{0k}$ and ${\mathbf{P}}_k$, the right amount of the
energy fluctuation can be produced by changing the momentum as
\begin{equation}
{\mathbf{P}}_k\rightarrow
(P_k-\delta p_k+x\Delta p_k){\mathbf{P}}_k/P_k,\label{eq:Pchange}
\end{equation}
where $x$ is a random number taken from the normal distribution with
$\langle x\rangle=0$ and $\langle x^2\rangle=1$, and
\begin{eqnarray}
\delta p_k&=&P_k-\sqrt{P_k^2-\hbar^2\nu},\label{eq:deltap}\\
\Delta p_k&=&\hbar\sqrt{\nu}\label{eq:Deltap}
\end{eqnarray}
Not only the fluctuation ($\Delta p_k$) is given but also the average
value of $P_k$ is decreased by $\delta p_k$ so that the energy
expectation value does not change by this random process.  It can be
introduced as a new term in the equation for ${\mathbf{\Xi}}_k$
(\ref{eq:EqOfXi}) which is put at the moment when the packet $k$ is
isolated ($N_k=0$) for the first time.  The momentum in the above
discussion should be understood as the relative momentum between the
emitted nucleon and the parent nucleus.  The total energy conservation
is achieved by adjusting other degrees of freedom of the parent
nucleus just in the same way for the fluctuation due to the wave
packet diffusion effect.

It should be emphasized here that the above prescription is taken when
the wave packet is emitted as a whole with the considerably high
momentum $P_k$, such as in the early stage of high energy collisions.
On the other hand, when a low energy nucleon evaporates from a
nucleus, what comes out of the nucleus is only a high momentum
component of the packet as we discussed in Refs.\
{}\cite{ONOf,ONOg,ONOh}. Therefore, the true momentum spread around
${\mathbf{P}}_k$ should be small and the above prescription should not
be taken in this case.  In order to continuously connect these two
extremes of high and low energies, Eqs.\ (\ref{eq:deltap}) and
(\ref{eq:Deltap}) are replaced by
\begin{eqnarray}
\delta p_k&=&P_k-\sqrt{\max(P_k^2-\hbar^2\nu,\ P_k^2/4)},\\
\Delta p_k&=&\sqrt{2P_k\delta p-\delta p^2},
\end{eqnarray}
and the momentum ${\mathbf{P}}_k$ is stochastically changed by Eq.\
(\ref{eq:Pchange}).

\subsection{Two-nucleon collision process}

The combination of the deterministic equation of motion and the
quantum branching process due to the wave packet diffusion effect is
essentially equivalent to the mean field theory, such as TDHF, for the
short time evolution of a channel wave function.  However, in medium
and high energy collisions, there should be the effect of the residual
interaction which brings a Slater determinant to a superposition of
many Slater determinants.  This effect is introduced as the stochastic
two-nucleon collision process.

In most molecular dynamics models {}\cite{AICHELIN,MARUb}, the
stochastic two-nucleon collision process has been introduced as the
process to cause such branchings.  In AMD {}\cite{ONOab}, two-nucleon
collisions are introduced by the use of the physical coordinates $W$
defined by Eq.\ (\ref{eq:PhysCoord}).  When the physical positions of
two nucleons get close, their physical momenta are changed randomly
according to the differential cross section in a similar way to QMD
{}\cite{AICHELIN,MARUb}. The energy-dependent collision cross section
may be modified due to the medium effect which can be taken into
account as the density dependence of the cross section.  The Pauli
blocking is automatically introduced because of the existence of the
Pauli-forbidden region in the physical coordinate space
{}\cite{ONOab}.

\section{Triple-Loop Approximation of AMD Hamiltonian}

Until recently, the application of AMD and AMD-V was limited to
relatively light systems with the total mass number $A<100$, because
the CPU time proportional to $A^4$ is necessary for the evaluation of
the interaction term in the AMD Hamiltonian,
\begin{equation}
{\mathcal{V}}=\frac{1}{2}\sum_{ijkl}
\langle\varphi_i\varphi_j|v|\varphi_k\varphi_l-\varphi_l\varphi_k\rangle
B^{-1}_{ki}B^{-1}_{lj},\label{eq:V2expectation}
\end{equation}
where $\varphi_i$ are the single particle wave functions in Eq.\
(\ref{eq:AMDWaveFunction}) and
$B_{ij}=\langle\varphi_i|\varphi_j\rangle$.  In order to apply AMD-V
to heavy systems such as $\gold+\gold$ collisions, we now introduce an
approximation for the AMD Hamiltonian.

First of all, by using the one-body Wigner function of Eq.\
(\ref{eq:AMDWignerFunction}), the two-body interaction term
(\ref{eq:V2expectation}) can be rewritten as a bilinear form of $f$,
\begin{eqnarray}
{\mathcal{V}}&=&f\cdot\hat v\cdot f \nonumber\\
  &\equiv& \int\frac{d{\mathbf{r}} d{\mathbf{p}} d{\mathbf{r}}' d{\mathbf{p}}'}
  {(2\pi\hbar)^6}
         f({\mathbf{r}},{\mathbf{p}})
         \hat v({\mathbf{r}},{\mathbf{p}};{\mathbf{r}}',{\mathbf{p}}')
         f({\mathbf{r}}',{\mathbf{p}}'),
\end{eqnarray}
where $\hat v$ includes the direct term and the exchange term,
\begin{eqnarray}
\hat v({\mathbf{r}},{\mathbf{p}};{\mathbf{r}}',{\mathbf{p}}')
&=&\frac{1}{2}v({\mathbf{r}}-{\mathbf{r}}') \nonumber\\
&&-\frac{1}{2}\delta({\mathbf{r}}-{\mathbf{r}}')
\int d{\mathbf{s}}\,
e^{-i({\mathbf{p}}-{\mathbf{p}}')\cdot{\mathbf{s}}/\hbar}v({\mathbf{s}}).
\end{eqnarray}
The spin and isospin degrees of freedom should be implicitly
understood.

The Wigner function is now approximated by a sum of $3A$ Gaussian
functions,
\begin{eqnarray}
f({\mathbf{u}})\approx f'''({\mathbf{u}})
&=&\sum_{p=1}^{3A}c_p f^{\mathrm{G}}_p({\mathbf{u}})
,\\
f^{{\mathrm{G}}}_p({\mathbf{u}})&\equiv&
 8e^{-2|{\mathbf{u}}-{\mathbf{w}}_p|^2}.
\end{eqnarray}
The centroids of $f^{\mathrm{G}}_p$ are chosen as
\begin{eqnarray}
{\mathbf{w}}_p=\left\{
\begin{array}{l@{\qquad}l}
{\mathbf{W}}_i                      & p=i\\
{\mathbf{Z}}_i+i({\mathbf{W}}_i-{\mathbf{Z}}_i) & p=A+i\\
{\mathbf{Z}}_i-i({\mathbf{W}}_i-{\mathbf{Z}}_i) & p=2A+i
\end{array}\right.\\
\mbox{for $i=1,\ldots,A$ and $p=1,\ldots,3A$,}\nonumber
\end{eqnarray}
so that the packets cover the important phase space region efficiently.
In order to get a good approximation, the coefficients $\{c_p\}$ are
determined by the condition
\begin{equation}
f^{\mathrm{G}}_p\cdot\hat{v}\cdot f'''
  = f^{\mathrm{G}}_p\cdot\hat{v}\cdot f\qquad\mbox{for $p=1,\ldots,3A$},
\label{eq:Determine_c}
\end{equation}
which means that the mean field $\hat{v}\cdot f$ averaged around the
phase space point ${\mathbf{w}}_p$ should not change when the exact Wigner
function $f$ is replaced by the approximated one $f'''$.  This
condition is just a linear equation system for $\{c_p\}$,
\begin{equation}
\sum_{q=1}^{3A}A_{pq}c_q=b_p,\label{eq:Determine_c_leq}
\end{equation}
with
\begin{eqnarray}
A_{pq}&=& f^{\mathrm{G}}_p\cdot\hat{v}\cdot f^{\mathrm{G}}_q,
\label{eq:DefA}\\
b_p   &=& f^{\mathrm{G}}_p\cdot\hat{v}\cdot f.
\label{eq:DefB}
\end{eqnarray}
The approximated value of ${\mathcal{V}}$ is then obtained by
\begin{equation}
{\mathcal{V}}\approx{\mathcal{V}}'''
\equiv f'''\cdot\hat{v}\cdot f'''=\sum_{pq}c_pA_{pq}c_q.
\end{equation}

In rare cases, the matrix $A_{pq}$ becomes close to a singular matrix.
Then the absolute values of $c_p$ are large, and $f'''$ oscillates
violently in the phase space so as to satisfy Eq.\
(\ref{eq:Determine_c}).  Since the intent of Eq.\
(\ref{eq:Determine_c}) is to reproduce $f$ by $f'''$ in the important
phase space region by requiring $f'''$ to be identical to $f$ around
the phase space points $\{{\mathbf{w}}_p\}$, the resultant $f'''$ should
be a smooth function for the consistency.  Therefore the above
equation (\ref{eq:Determine_c_leq}) is slightly modified to
\begin{equation}
\sum_q(A^2+\epsilon^2)_{pq}c_q=\sum_q A_{pq}b_q+\epsilon^2 c_p^0,
\label{eq:Determine_c_leq2}
\end{equation}
where $\epsilon$ is a small parameter, and
\begin{equation}
c_p^0=\left\{
\begin{array}{ll}
1&\mbox{for $p=1,\ldots,A$}\\
0&\mbox{for $p=A+1,\ldots,3A$}
\end{array}\right. .
\end{equation}
This modified equation is the condition to minimize the quantity
\begin{equation}
\sum_p(f_p^{\mathrm{G}}\cdot\hat{v}\cdot f'''
       -f_p^{\mathrm{G}}\cdot\hat{v}\cdot f)^2
+\epsilon^2\sum_p(c_p-c_p^0)^2
\end{equation}
with respect to the coefficients $\{c_p\}$, so that we can avoid the
situation where $c_p$ deviates from the normal value $c_p^0$ very
much.

It can be easily seen that the approximated interaction
${\mathcal{V}}'''$ can be evaluated with the CPU time proportional to
$A^3$ which is necessary for evaluating $\{b_p\}$ by Eq.\
(\ref{eq:DefB}) and for solving Eq.\ (\ref{eq:Determine_c_leq2}). We
can also use a similar approach to approximate the derivatives of
${\mathcal{V}}$ with respect to the coordinates $Z$.  The required CPU
time is also proportional to $A^3$.

\iffloat
\begin{figure*}
\noindent
\begin{minipage}[t]{0.45\textwidth}
\begin{center}
\includegraphics[width=0.9\textwidth]{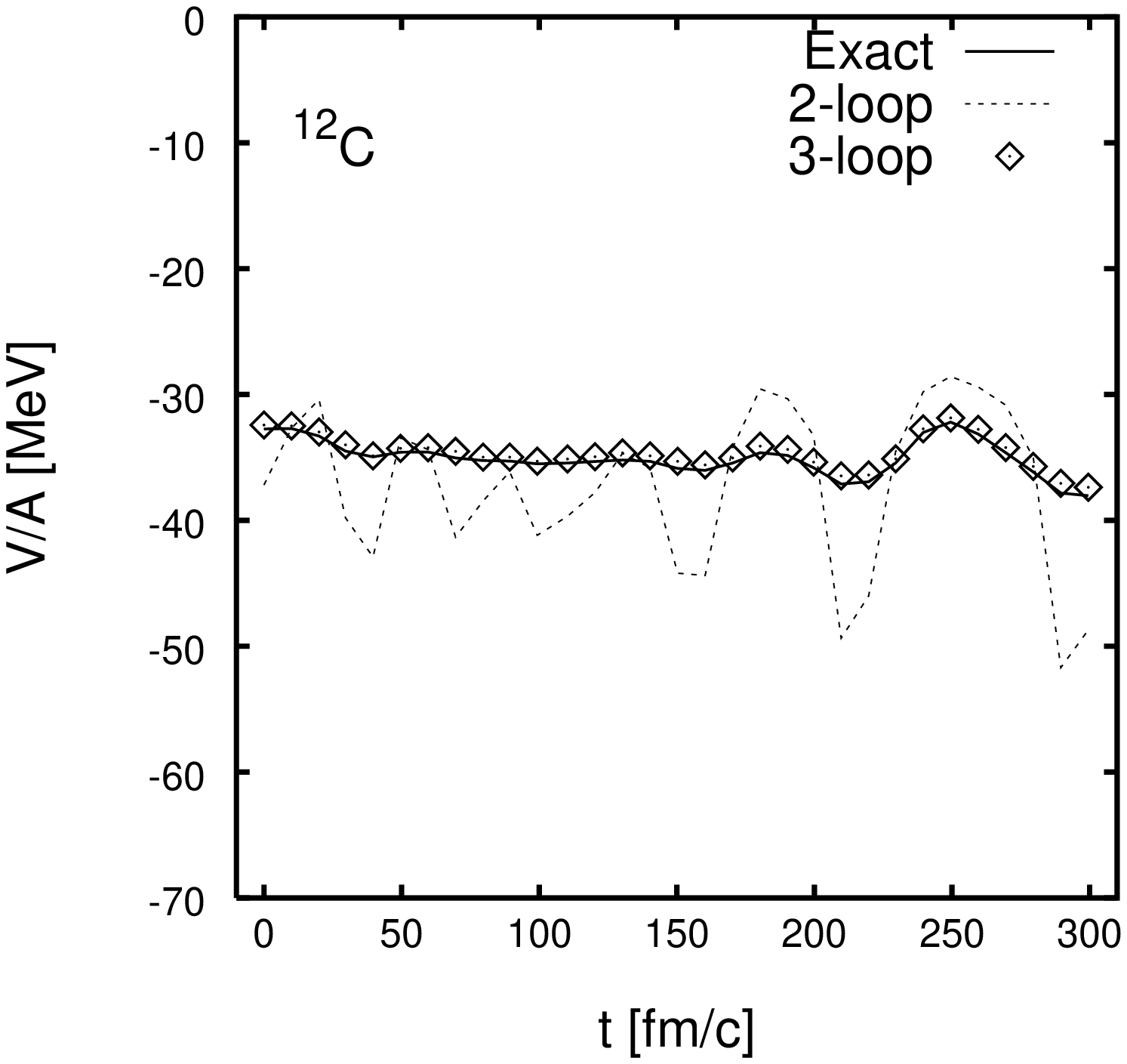}
\end{center}
\vspace{-0.15\textwidth}
\begin{center}
\includegraphics[width=0.9\textwidth]{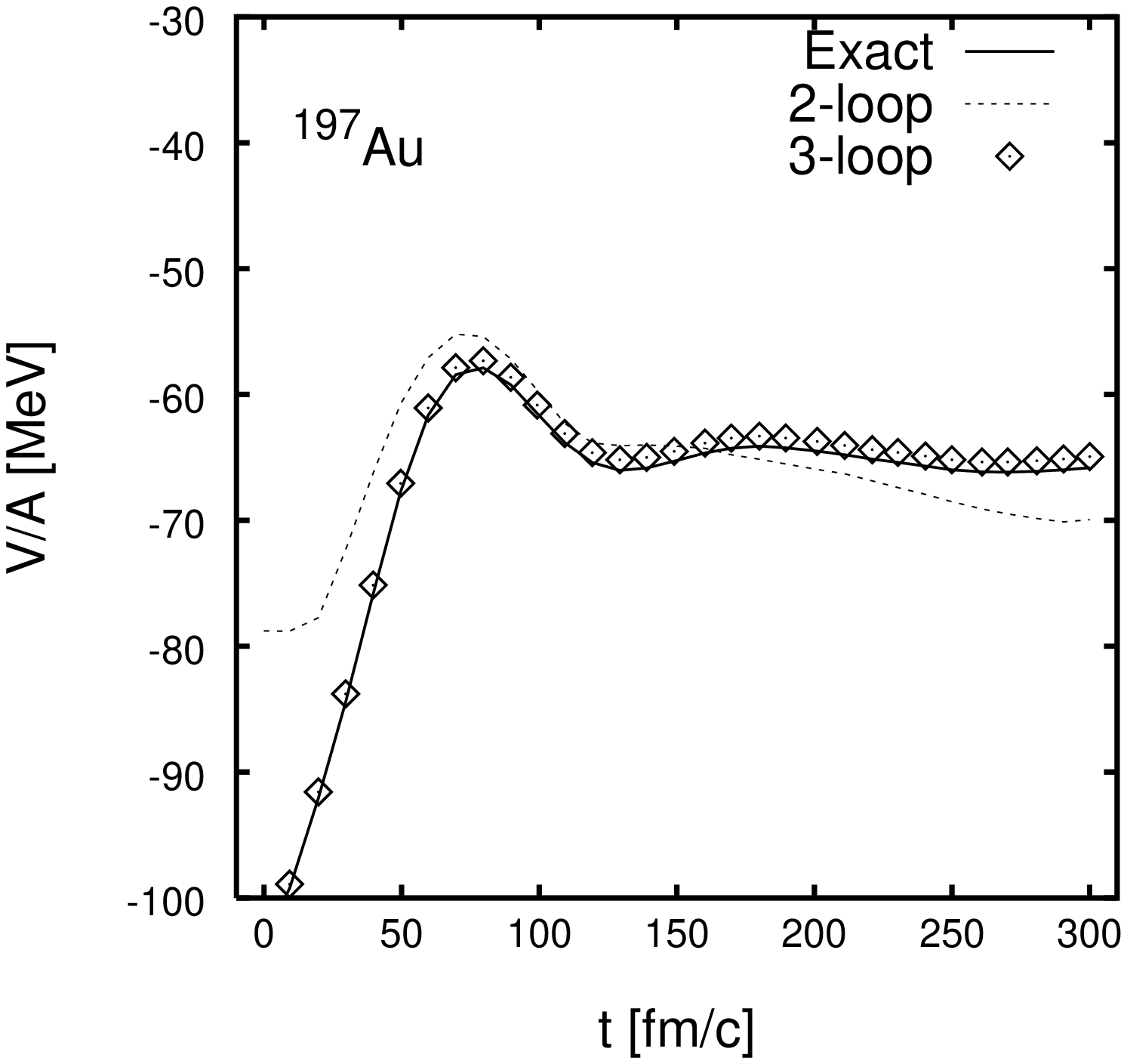}
\end{center}
\caption{\label{fig:coolcheck}
Tests of the triple-loop approximation (diamonds) along the slow
frictional cooling path for $\carbon$ and $\gold$ nuclei, compared
with the exact values (solid line).  The results of the QMD-like
approximation (dotted line) are also shown.  The expectation value of
the two-body part of the Gogny force is shown as a function of time.}
\end{minipage}
\hspace{0.08\textwidth}
\begin{minipage}[t]{0.45\textwidth}
\begin{center}
\includegraphics[width=0.9\textwidth]{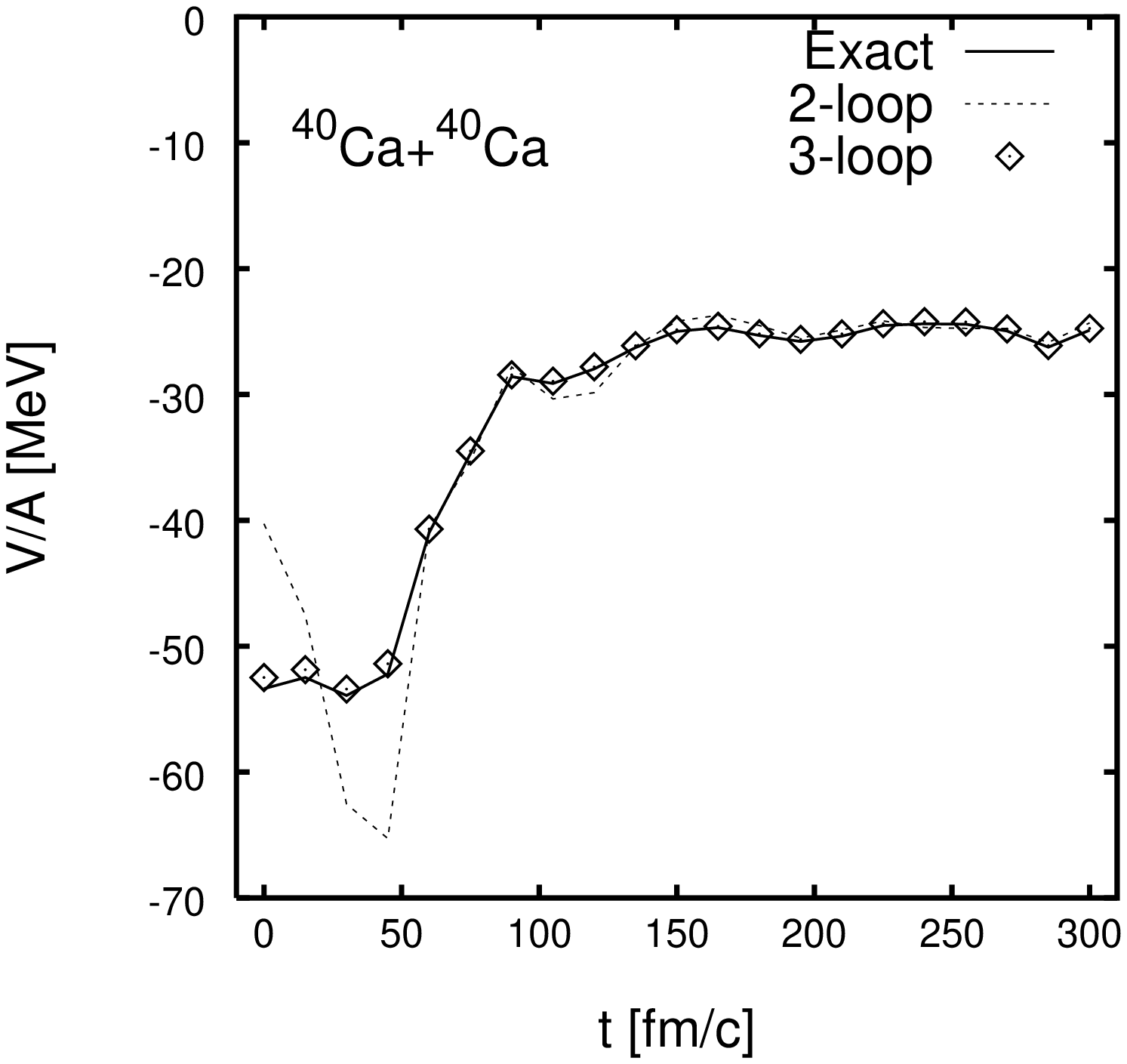}
\end{center}
\vspace{-0.15\textwidth}
\begin{center}
\includegraphics[width=0.9\textwidth]{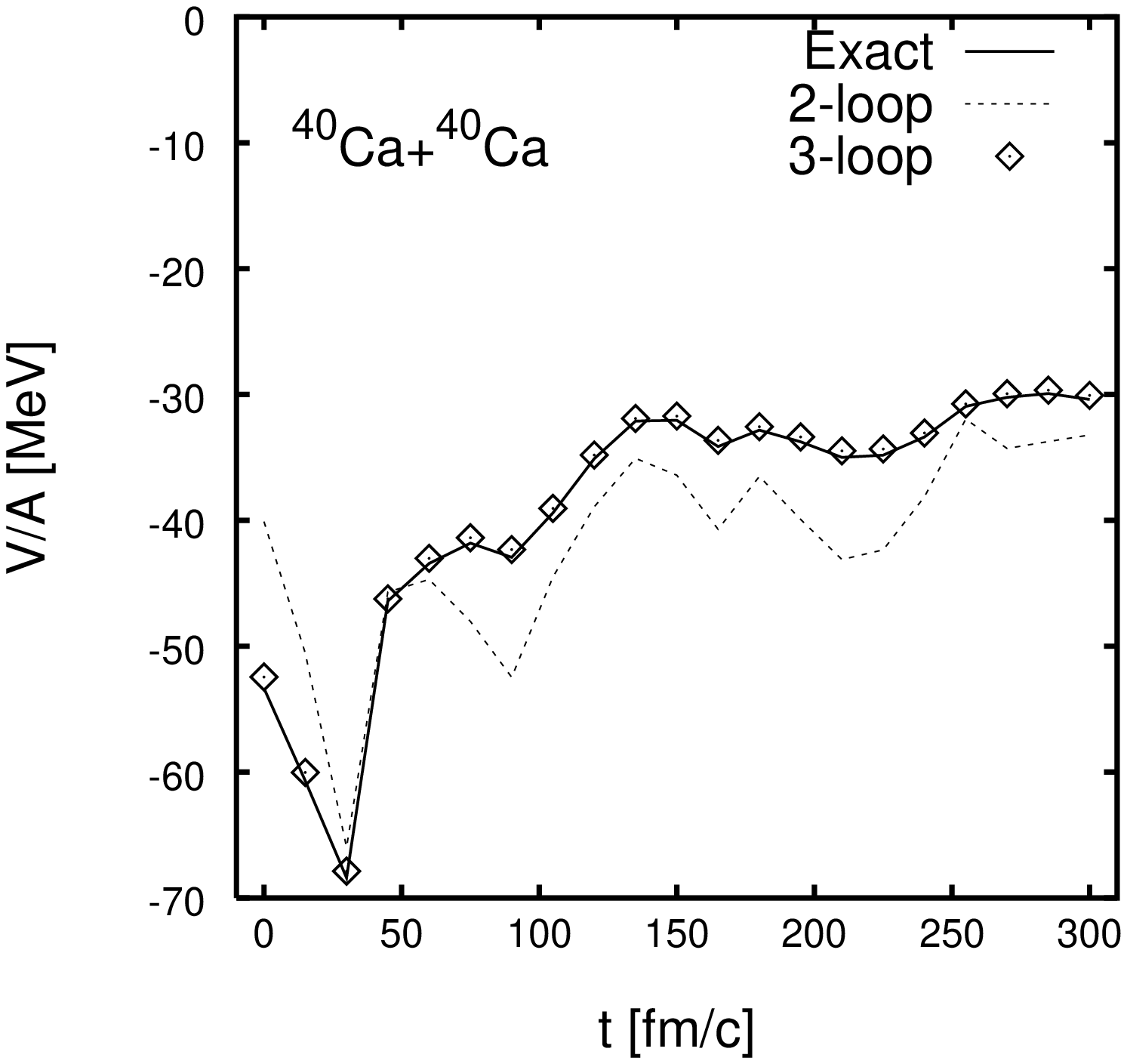}
\end{center}
\caption{\label{fig:anacheck}
The same as Fig.\ {}\ref{fig:coolcheck}, but for the tests along the
dynamics of $\calcium+\calcium$ collisions at 35 MeV/nucleon.}
\end{minipage}
\end{figure*}
\fi

The above formalism can be applied for the density-dependent zero-range
force with a little extension.  The forces like the Gogny force and
the Skyrme force have the density-dependent term
\begin{equation}
\hat v({\mathbf{r}},{\mathbf{p}};{\mathbf{r}}',{\mathbf{p}}')
=v_0 [\rho({\mathbf{r}})]^\sigma\delta({\mathbf{r}}-{\mathbf{r}}').
\label{eq:VhatDD}
\end{equation}
The coefficients $\{c_p\}$ are determined in the same way as above by
using Eqs.\ (\ref{eq:DefA}), (\ref{eq:DefB}) and
(\ref{eq:Determine_c_leq2}) but by replacing the density
$\rho({\mathbf{r}})$ in Eq.\ (\ref{eq:VhatDD}) by a constant $\rho_0$.
The result $\{c_p\}$ does not depend on the value of $\rho_0$.  Then
the approximated value of ${\mathcal{V}}$ is obtained by
\begin{equation}
{\mathcal{V}}\approx{\mathcal{V}}'''
=\sum_{pq}
\biggl(\frac{\tilde{\rho}_p\tilde{\rho}_q}{\rho_0^2}\biggr)^{\sigma/2}
c_p A_{pq}c_q,
\end{equation}
where $\tilde{\rho}_p$ is a smoothed density around the point
$\Re{\mathbf{w}}_p$ defined by
\begin{equation}
\tilde{\rho}_p=
\biggl(\frac{\mu\nu}{\pi}\biggr)^{3/2}
\sum_q c_q e^{-\mu|\Re{\mathbf{w}}_p-\Re{\mathbf{w}}_q|^2}.
\end{equation}
The parameter $\mu=\frac{4}{3}$ is chosen so as to give a good
approximation.

This triple-loop approximation is tested under various
circumstances. Figure {}\ref{fig:coolcheck} shows the test along
the slow frictional cooling path [Eq.\ (\ref{eq:EqOfCooling}) with
$\lambda=1$ and $\mu=-0.25$] for two nuclei $\carbon$ and $\gold$.
The randomly excited initial nuclei at $t=0$ fm/$c$ are cooled down to
the ground states at $t\sim 300$ fm/$c$.  The exact expectation value
${\mathcal{V}}/A$ of the density-independent two-body part of the Gogny
force is shown by a solid line, while the approximated value
${\mathcal{V}}'''/A$ with the triple-loop approximation is shown by a
diamond for each $t$.  The dotted line shows the result of the
QMD-like approximation where the expectation value is evaluated by
using the approximated Wigner function of Eq.\
(\ref{eq:WignerFunctionQMD}),
\begin{equation}
{\mathcal{V}}''=\sum_{p=1}^A\sum_{q=1}^A A_{pq}.
\end{equation}
Compared to the too bad result of the QMD-like approximation, the
triple-loop approximation always gives a good result within the error
of about 1 MeV/nucleon.  Figure {}\ref{fig:anacheck} shows the similar
information for two events of $\calcium+\calcium$ collisions at 35
MeV/nucleon.  The event of the upper part is a peripheral collision
and the event of the lower part is a central collision.  The
triple-loop approximation again gives a sufficiently good result of
the error within 1 MeV/nucleon.  We have also found that the
triple-loop approximation for the density-dependent force has the
precision similar to the density-independent force.

\section{An Application to A\lowercase{u} + A\lowercase{u} Collisions}

\iffloat
\begin{figure*}
\noindent
\includegraphics[width=\textwidth]{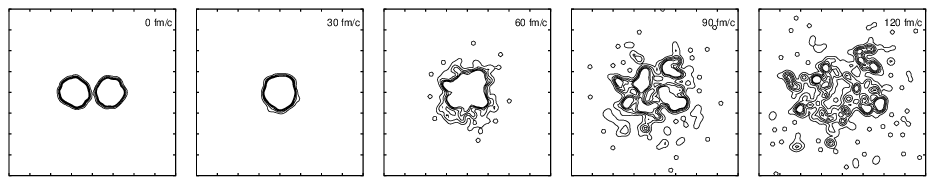}
\includegraphics[width=\textwidth]{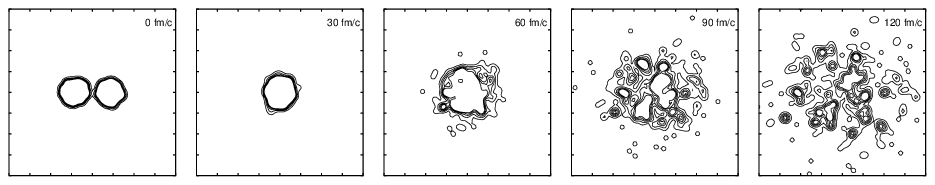}
\caption{\label{fig:AuAu-anim}
Examples of the time evolution of the density projected onto the
reaction plane from $t=0$ fm/$c$ to $t=120$ fm/$c$ for central
$\gold+\gold$ collisions at 150 MeV/nucleon. The size of the shown
area is 80 fm $\times$ 80 fm.}
\end{figure*}
\fi

The $\gold+\gold$ collisions are calculated by AMD-V for the incident
energy 150 MeV/nucleon and the impact parameter $0<b<1$ fm.  One of
the interesting aspects in this reaction is the copious formation of
the intermediate mass fragments (IMFs) with $Z\ge3$ from a strongly
expanding system as observed in the experiment of Ref.\
{}\cite{REISDORF}.  On the other hand, only few IMFs are produced in
the dynamical QMD calculation {}\cite{REISDORF}.  Therefore it is an
important theoretical problem to find out how the copious fragment
formation is understood in the dynamical framework.

In the AMD-V calculation presented here, the Gogny force
{}\cite{GOGNY} is adopted as the effective interaction.  It
corresponds to the soft equation of state with the incompressibility
$K=228$ MeV and the appropriate momentum dependence of the mean field.
The expectation value of the Hamiltonian is evaluated by using the
triple-loop approximation described in the previous section.  The
ground state of the $\gold$ nucleus is obtained by the frictional
cooling method, and it has the reasonable binding energy $E/A=7.4$ MeV
and the root mean square radius $\langle r^2\rangle^{1/2}=5.5$ fm,
while the experimental data are 7.9 MeV and 5.3 fm, respectively.  The
adopted two nucleon collision cross section and the angular
distribution are the same as those of Ref.\ {}\cite{ONOd}.  Around the
two-nucleon collision energy $E_{\mathrm{NN}}\sim150$ MeV, which is
important in the present reaction, the $pp$ and $nn$ cross section is
the same as the free cross section (25 mb).  The $pn$ cross section is
the same as the free cross section (40 mb) at zero-density, but it is
reduced to about 30 mb for $\rho>\rho_0$ as the medium effect.

The produced fragments in the dynamical AMD-V calculation are
generally excited and their decay is calculated by a statistical
model.  At every 15 fm/$c$ in the dynamical AMD-V calculation, the
fragments are identified by linking the two-nucleon pairs with
$|{\mathbf{Z}}_i-{\mathbf{Z}}_j|/\sqrt{\nu}< 5$ fm.  With this
condition, the identified fragments are well separated spatially in
most cases.  A fragment with the mass number $5\le A< A_{\mathrm{cr}}$
is thrown to the statistical decay code directly if its mass number
before 15 fm/$c$ was also $5\le A< A_{\mathrm{cr}}$.  Namely, the
statistical decay of each primordial fragment is calculated when the
waiting time $t_{\mathrm{wait}}$ has passed since its mass first
became smaller than $A_{\mathrm{cr}}$.  The parameters are
$t_{\mathrm{wait}}=22.5$ fm/$c$ on the average, and $A_{\mathrm{cr}}$
is chosen to be 25.  However, the dependence of the results on these
parameters is found to be small.  When we take $A_{\mathrm{cr}}=20$ or
30, or $t_{\mathrm{wait}}=37.5$ fm/$c$ on the average, the change of
the IMF multiplicity is a few percent at most.  The adopted
statistical decay code {}\cite{MARUb} is based on the sequential
binary decay model by P\"uhlhofer {}\cite{PUHLHOFER}, but it also
takes account of the emission of composite particles not only in their
ground states but also in their excited states with the excitation
energy $E^*\le 40$ MeV.

\iffloat
\begin{figure}
\begin{center}
\includegraphics[width=0.5\textwidth]{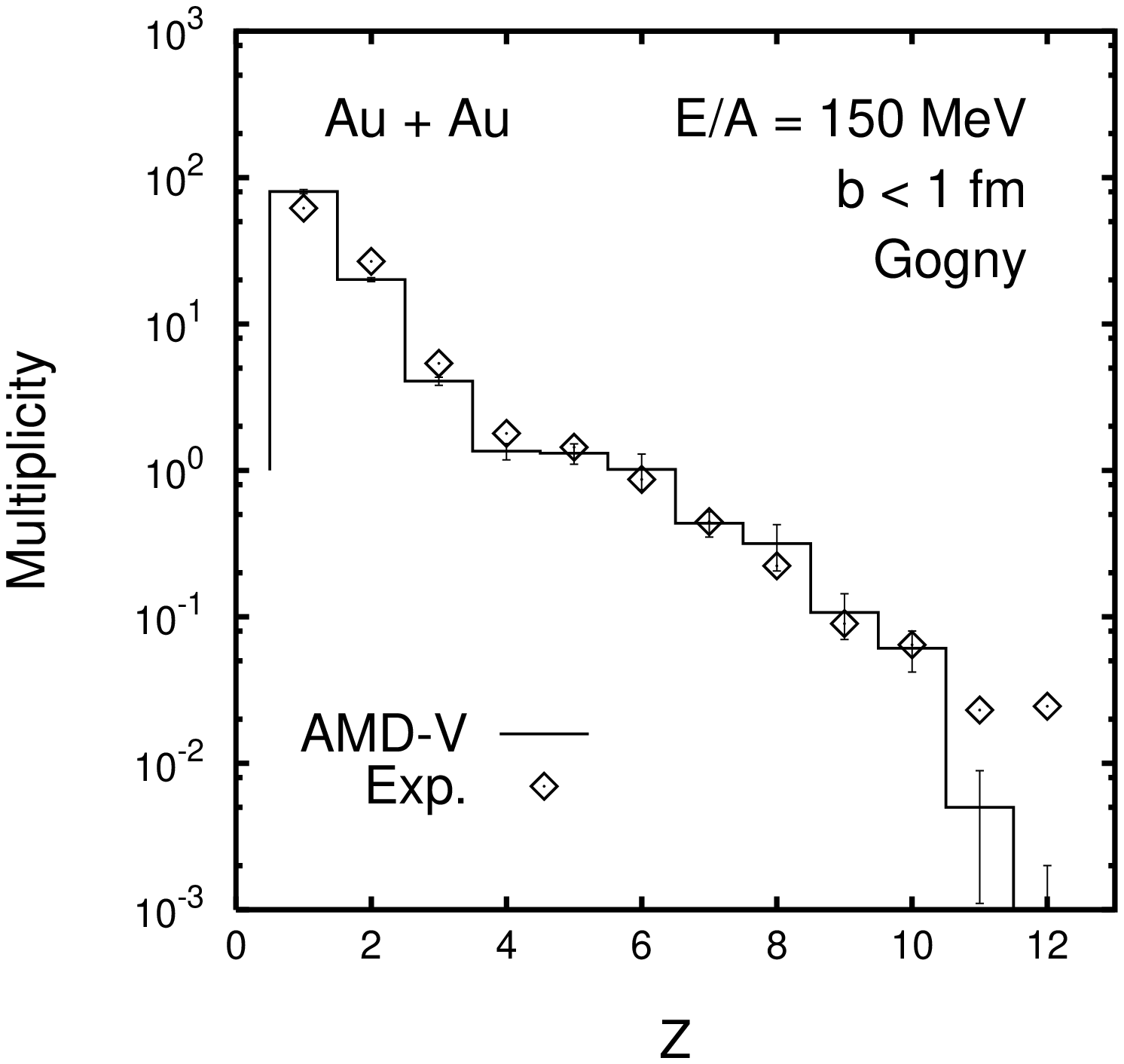}
\end{center}
\caption{\label{fig:AuAu-zmulti-E150}
Calculated charge distribution (histogram) in central $\gold+\gold$
collisions at 150 MeV/nucleon, compared with the
experimental data (diamonds) of Ref. {}\protect\cite{REISDORF}.  The
error bars show the estimated statistical error of the calculated
results.}
\end{figure}
\fi

Figure {}\ref{fig:AuAu-anim} shows the time evolution of the density
projected to the reaction plane for two events.  The total system is
once compressed and then expands rather rapidly.  From the expanding
matter, a lot of IMFs are produced. The multiplicity of the primordial
IMFs is about 16, and about half of them are to disappear by the
statistical decay.  Although the stopping seems to be strong and the
expansion is almost isotropic, the mixture of the projectile and the
target is not complete.  More wave packets of the projectile origin
come out to the forward direction than to the backward direction.

\iffloat
\begin{table}
\caption{\label{table:AuAu-multi-E150}
Multiplicities of various particles in central $\gold+\gold$
collisions at 150 MeV/nucleon.
}
\begin{center}
\begin{minipage}{0.45\textwidth}
\begin{tabular}{ldd}
 & Experiment\protect\cite{REISDORF} & AMD-V\\
\hline
neutron             &  92.6 &   120.6 \\
proton              &  26.1 &    56.8\\
deuteron            &  18.6 &    14.7\\
triton              &  17.2 &     8.8\\
${}^3\mathrm{He}$   &   5.7 &     2.3\\
${}^4\mathrm{He}$   &  21.0 &    16.3\\
IMF                 &  10.4 &     8.7
\end{tabular}
\end{minipage}
\end{center}
\end{table}
\fi

In Fig.\ {}\ref{fig:AuAu-zmulti-E150}, the calculated charge
distribution is compared with the experimental data {}\cite{REISDORF}.
The calculated result, shown by the solid histogram, reproduces the
data very well at least in the logarithmic scale.  The multiplicities
of various particles are compared to the data in Table
{}\ref{table:AuAu-multi-E150}.  The large IMF multiplicity of the
experimental data $M_{\mathrm{IMF}}=10.4$ is almost reproduced by the
calculated value $M_{\mathrm{IMF}}=8.7$, though it is slightly smaller
than the data.  This underestimation is due to the underestimation of
the Be and Li multiplicities.  We also notice in Table
{}\ref{table:AuAu-multi-E150} that the calculated multiplicities of
light particles with $2\le A\le4$ are too small and the nucleon
multiplicity is too large.

It is useful to consider the gas and liquid parts separately.  Here
the gas part is composed the freely moving nucleons and light
particles which are usually emitted after hard two-nucleon collisions,
and the liquid part is composed of the IMFs which is still bound well
by the mean field.  In order to properly describe the fragment
formation, the separation of the mass and the energy to the gas and
liquid parts is essential as well as the dynamics of the liquid part
itself.  The good reproduction of the IMF multiplicity and the charge
distribution for $Z\gtrsim 5$ suggests that AMD-V describes these
aspects very well.  On the other hand, the failure in the light
particle multiplicities can be regarded as a problem in the dynamics
of the gas part, namely the coalescence of particles in the gas part.

The light particle and IMF multiplicities are much better reproduced
when AMD-V is augmented by the coalescence of nucleons and light
particles, as will be shown in another paper.  In this paper, we just
mention why the coalescence is not properly treated in AMD-V and
should be added to AMD-V as an augmenting process.  In the medium and
high energy collisions like the present reaction, a lot of nucleons
are emitted.  Even though these nucleons have almost no correlations
among them after hard two-nucleon collisions, a pair of a proton and a
neutron can form a deuteron when these two nucleons are accidentally
close to each other in the phase space.  In order to correctly predict
the probability of the coalescence of uncorrelated nucleons, it is
necessary for AMD-V to have the correct phase space volume for the
bound deuteron state.  The phase space volume for the wave packet
centroids is important because the dynamics is governed by the
apparently classical equation of motion for the centroids.  Since the
deuteron is a loosely bound system with a single bound state, the
phase space volume in AMD is much smaller than the correct quantum
phase space $(2\pi\hbar)^3$.  The deuteron and nucleon yields should
therefore be underestimated and overestimated, respectively.  A large
part of tritons and ${}^{3}\mathrm{He}$ may also be produced by the
coalescence mechanism of three nucleons (or a nucleon and a deuteron),
and therefore the present calculation naturally underestimates their
multiplicities.  Furthermore, we should note that the intrinsic bound
states of Li and Be isotopes have the cluster structure of light
composite particles such as $\alpha$, $t$ and ${}^{3}\mathrm{He}$,
with the small binding energies between them.  The bound phase space
volume in AMD is likely to be smaller than the correct quantum phase
space, and it is natural that AMD-V underestimates the coalescence of
the light composite particles to produce Li and Be isotopes directly.

\section{Summary}

The quantum branching processes are essential in the molecular
dynamics models in order to properly describe the multichannel
reactions such as the multifragmentation in heavy ion collisions.  In
addition to the two-nucleon collision process which has been
recognized as an important process, AMD-V takes account of the wave
packet diffusion as a stochastic branching process rather than as the
shape change of the single particle wave packet like in TDHF or the
Vlasov equation.  AMD-V and the Vlasov equation are equivalent with
respect to the infinitesimal time evolution of a single Slater
determinant without two-nucleon collisions, except that AMD-V ignores
the interference among the channels and avoids the spurious channel
correlations.

In this paper, we reformulated AMD-V in two points so that it is
applicable even to heavy systems such as $\gold+\gold$ collisions.
Fist, the fluctuation due to the wave packet diffusion was formulated
as a stochastic term in the equation of motion for the wave packet
centroids.  A small Gaussian fluctuation is given to each packet at
every time step, instead of a big displacement once in a while in the
previous framework.  This reformulation decreases the numerical labor
because it simplifies the energy conservation procedure.  Second, a
new triple-loop approximation was introduced for the expectation value
of the Hamiltonian with respect to the AMD wave function.  With this
triple-loop approximation, the expectation value can be evaluated with
the numerical operations proportional to $A^3$ instead of $A^4$ in the
exact calculation, where $A$ is the mass number of the total system.
The error of this approximation is about 1 MeV/nucleon at most, and
therefore it is useful for the study of heavy ion collisions.

The reformulated AMD-V was applied to $\gold+\gold$ central collisions
at 150 MeV/nucleon.  We adopted the Gogny force as the effective
interaction.  The calculation reproduces the qualitative feature of
the experimental data that a lot of fragments are produced from the
radially expanding system.  The large IMF multiplicity was almost
reproduced by AMD-V quantitatively.  This result therefore suggests
that AMD-V works well for the aspects related to the fragment
formation, such as the large energy carried out by light particles,
the collective expansion, and the appearance of the cluster
correlation in the expanding system.  However, we found that the
nucleon multiplicity is strongly overestimated and the other light
particle multiplicities are underestimated.  This should be due to the
problem of AMD-V in the description of the coalescence of nucleons and
light particles which is beyond the scope of the current version of
AMD-V or other usual mean field theories.  We will show in another
paper how we can incorporate the coalescence to AMD-V and that the
coalescence process improves the reproduction of the data, including
the reactions with higher energy.  Furthermore, it is also an
interesting subject in progress to study the fragment formation in
relation to the equation of state of the nuclear matter in high and
low density, and also with isospin asymmetry.

\acknowledgments

The author would like to thank Prof.\ H. Horiuchi for useful
discussions and encouragements.  The numerical calculation was
performed by using NEC SX4 of RCNP, Osaka University, and Fujitsu
VPP-500 of KEK.

\iffloat
\end{document}
\fi

\begin{table}
\caption{\label{table:AuAu-multi-E150}
Multiplicities of various particles in central $\gold+\gold$
collisions at 150 MeV/nucleon.
}
\begin{center}
\begin{minipage}{0.45\textwidth}
\begin{tabular}{ldd}
 & Experiment\protect\cite{REISDORF} & AMD-V\\
\hline
neutron             &  92.6 &   120.6 \\
proton              &  26.1 &    56.8\\
deuteron            &  18.6 &    14.7\\
triton              &  17.2 &     8.8\\
${}^3\mathrm{He}$   &   5.7 &     2.3\\
${}^4\mathrm{He}$   &  21.0 &    16.3\\
IMF                 &  10.4 &     8.7
\end{tabular}
\end{minipage}
\end{center}
\end{table}

\begin{figure}
\begin{center}
\begin{minipage}{0.45\textwidth}
\begin{center}
\includegraphics[width=\textwidth]{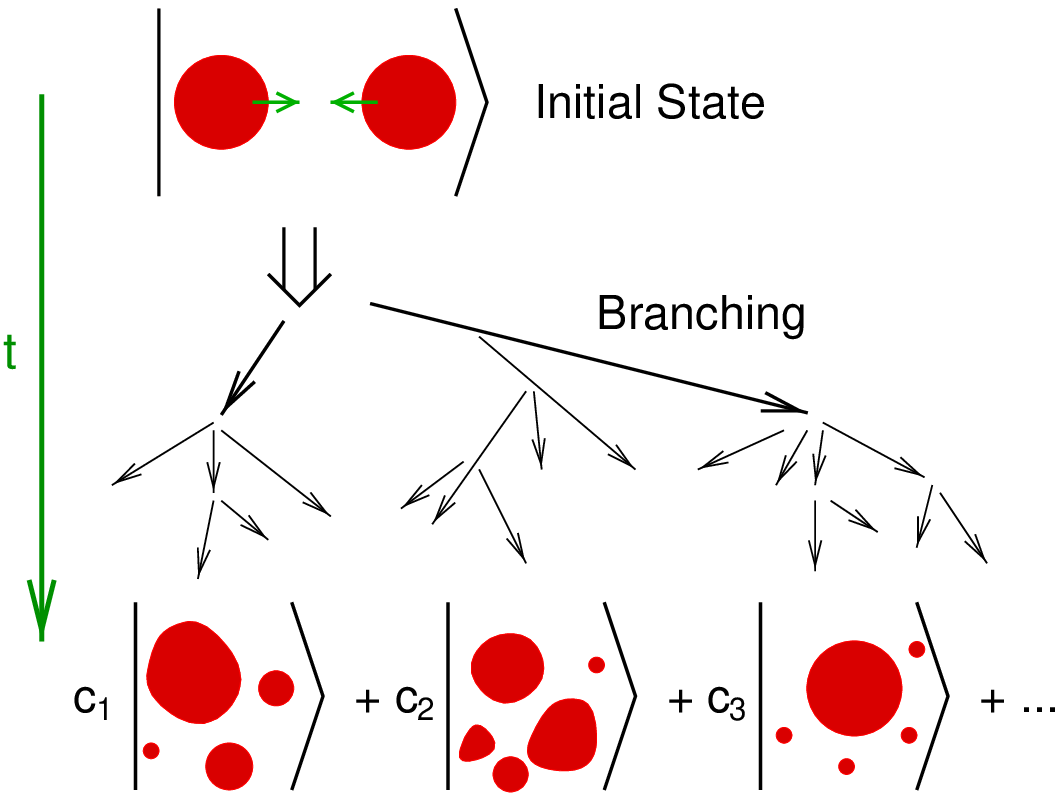}
\end{center}
\end{minipage}
\end{center}
\caption{\label{fig:multichannel}
A schematic picture of the quantum branching processes for
multichannel reactions.}
\end{figure}
\newpage

\begin{figure}
\noindent
\begin{center}
\begin{minipage}[t]{0.45\textwidth}
\includegraphics[width=0.9\textwidth]{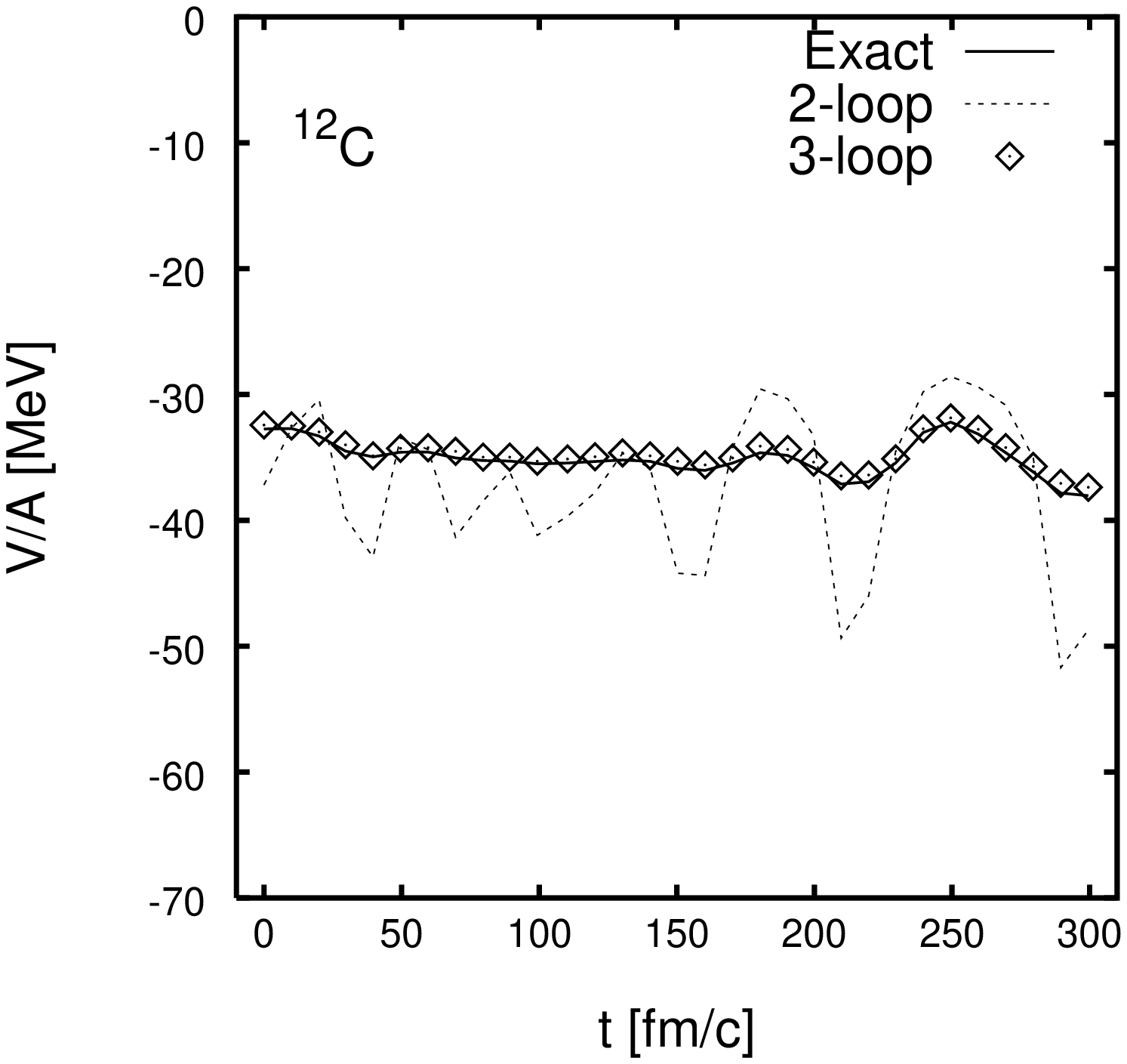}
\includegraphics[width=0.9\textwidth]{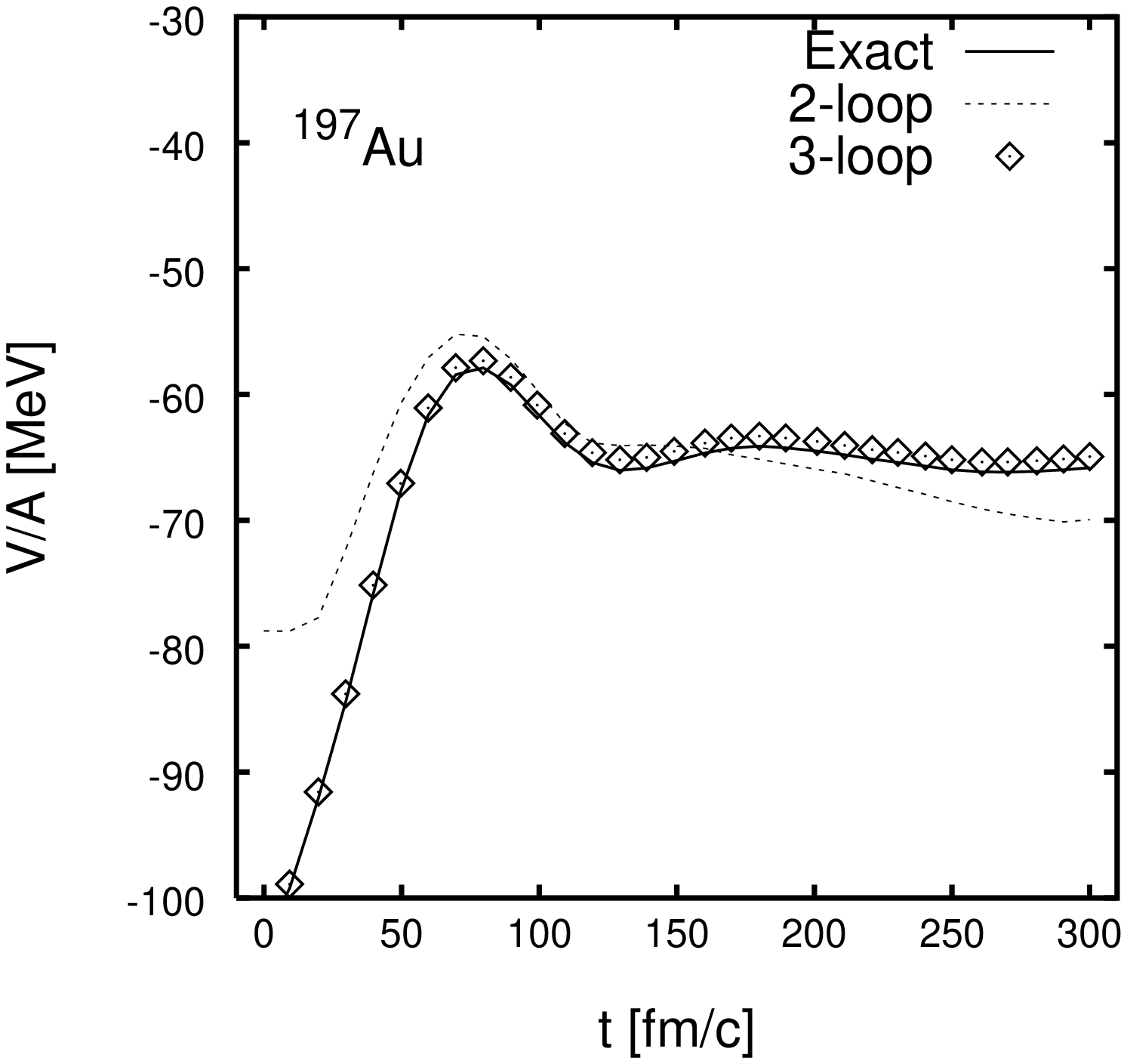}
\end{minipage}
\end{center}
\caption{\label{fig:coolcheck}
Tests of the triple-loop approximation (diamonds) along the slow
frictional cooling path for $\carbon$ and $\gold$ nuclei, compared
with the exact values (solid line).  The results of the QMD-like
approximation (dotted line) are also shown.  The expectation value of
the two-body part of the Gogny force is shown as a function of time.}
\end{figure}
\newpage

\begin{figure}
\begin{center}
\begin{minipage}[t]{0.45\textwidth}
\includegraphics[width=0.9\textwidth]{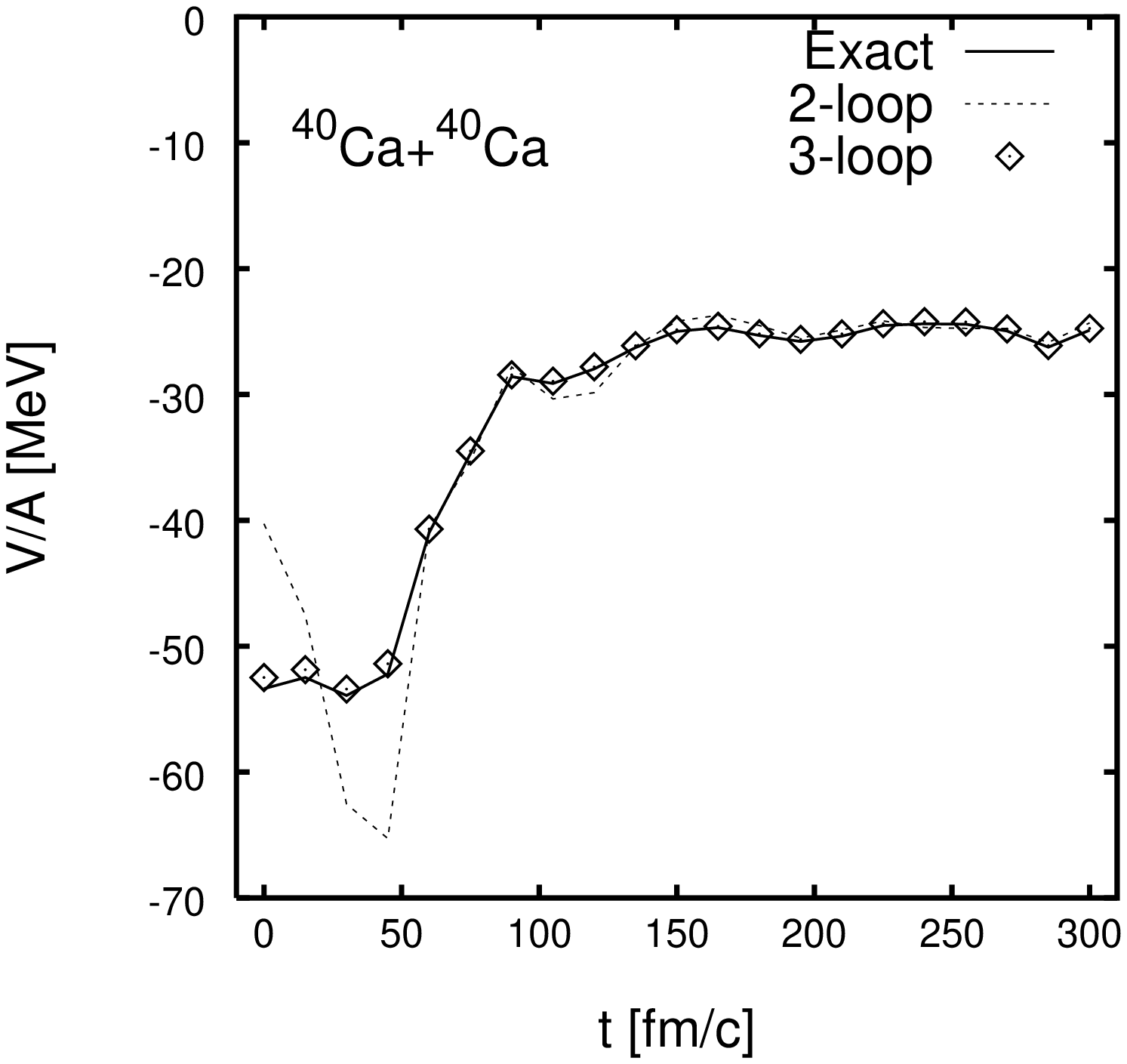}
\includegraphics[width=0.9\textwidth]{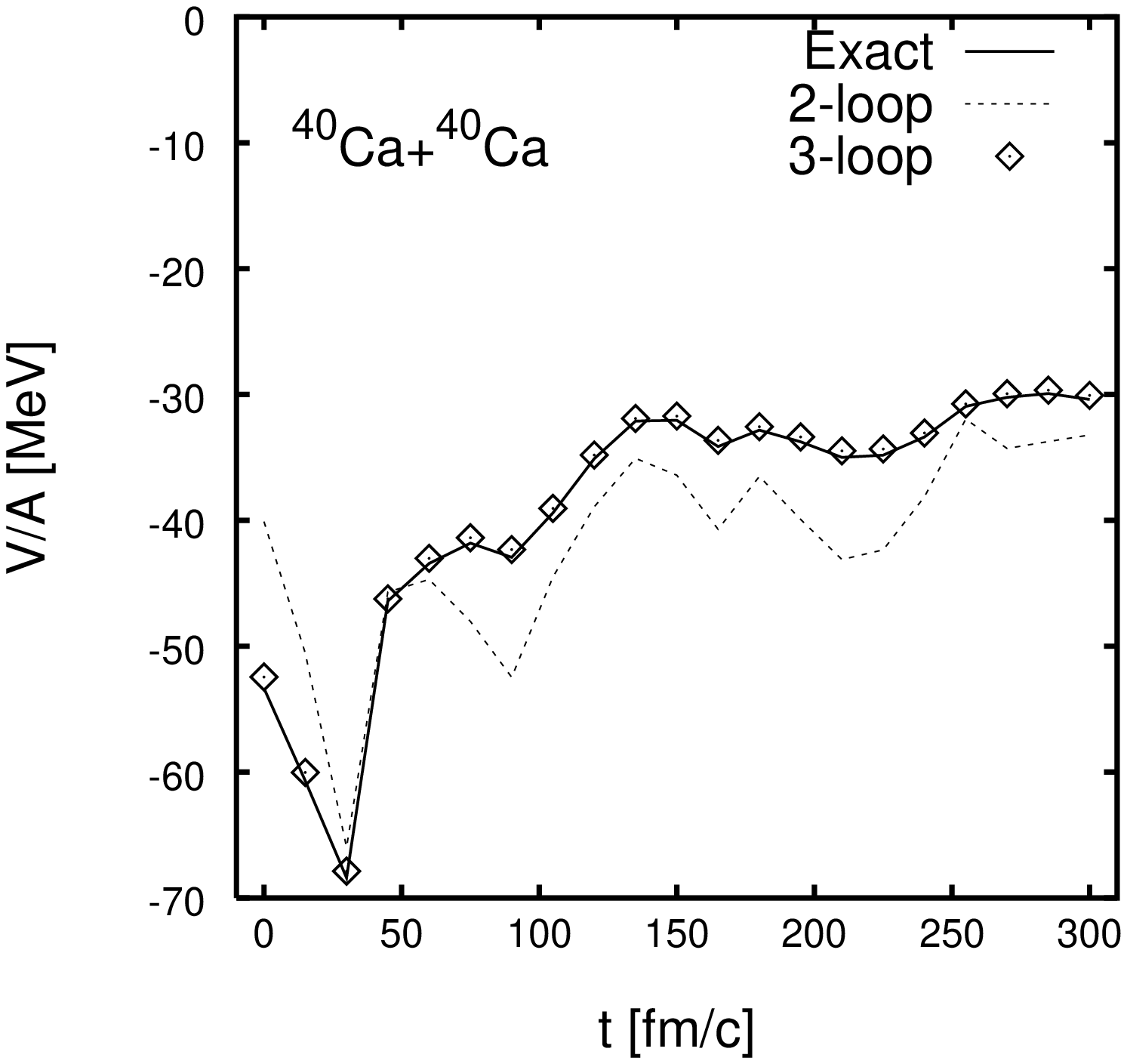}
\end{minipage}
\end{center}
\caption{\label{fig:anacheck}
The same as Fig.\ {}\ref{fig:coolcheck}, but for the tests along the
dynamics of $\calcium+\calcium$ collisions at 35 MeV/nucleon.}
\end{figure}
\newpage

\begin{figure}
\noindent
\includegraphics[width=\textwidth]{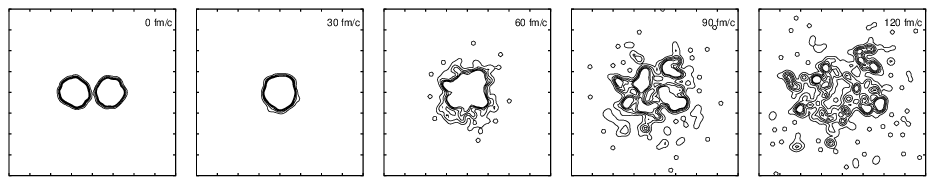}
\includegraphics[width=\textwidth]{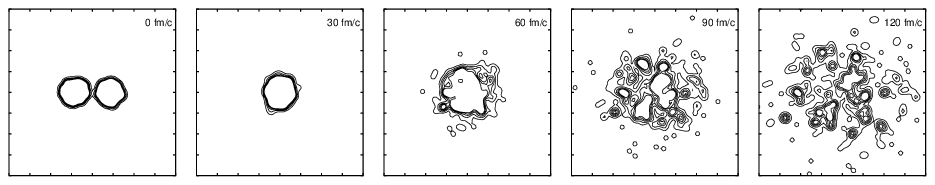}
\caption{\label{fig:AuAu-anim}
Examples of the time evolution of the density projected onto the
reaction plane from $t=0$ fm/$c$ to $t=120$ fm/$c$ for central
$\gold+\gold$ collisions at 150 MeV/nucleon. The size of the shown
area is 80 fm $\times$ 80 fm.}
\end{figure}
\newpage

\begin{figure}
\begin{center}
\includegraphics[width=0.5\textwidth]{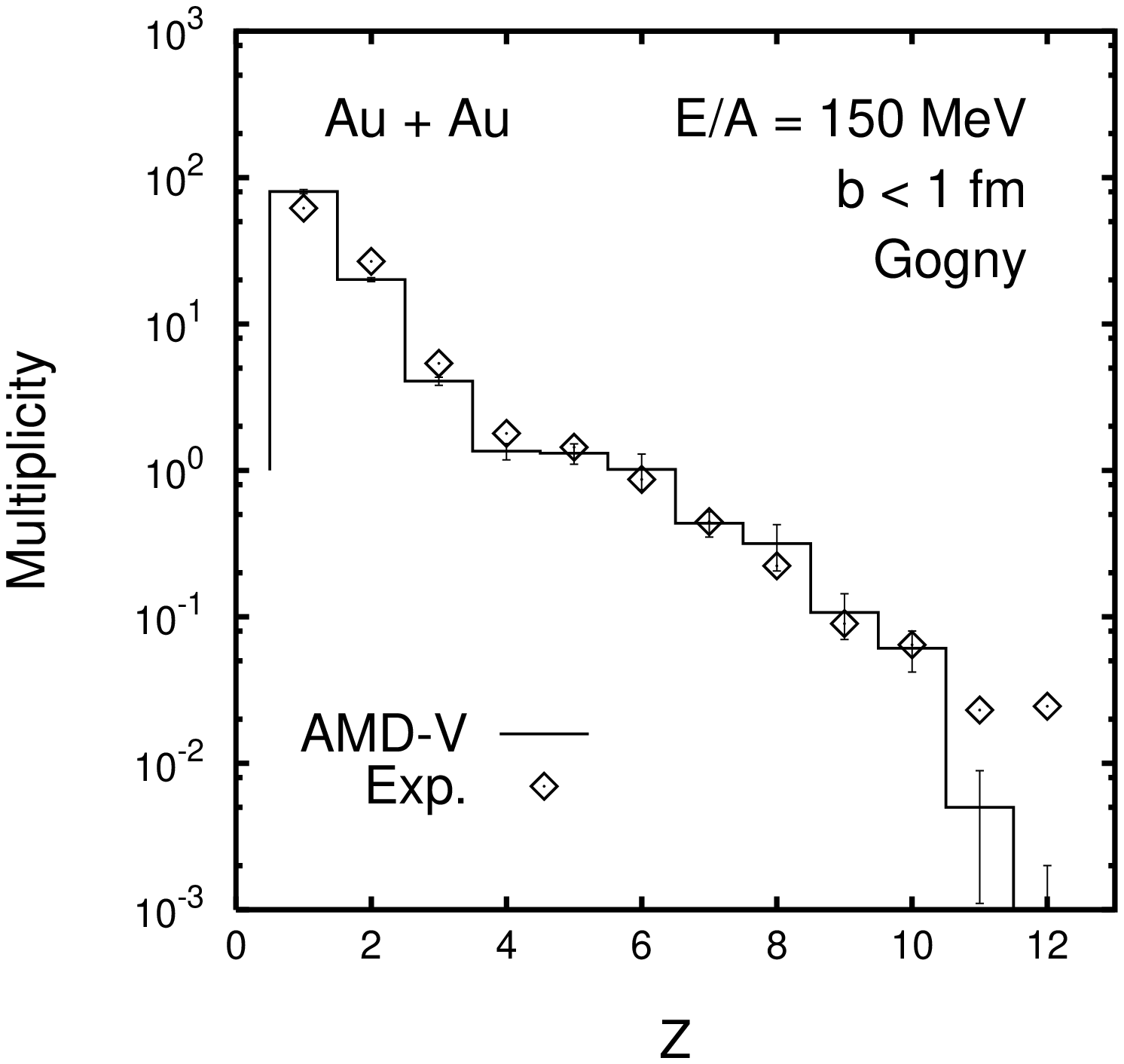}
\end{center}
\caption{\label{fig:AuAu-zmulti-E150}
Calculated charge distribution (histogram) in central $\gold+\gold$
collisions at 150 MeV/nucleon, compared with the
experimental data (diamonds) of Ref. {}\protect\cite{REISDORF}.  The
error bars show the estimated statistical error of the calculated
results.}
\end{figure}

\end{document}